\documentclass[10pt,onecolumn,aps,prd,preprintnumbers,showpacs,superscriptaddress,nofootinbib,amsmath,amssymb,floats,floatfix,showkeys,notitlepage,longbibliography]{revtex4-2}
\usepackage{comment}
\usepackage{lipsum}
\usepackage{graphicx}
\usepackage{subfigure}
\usepackage{palatino}
\usepackage{sans}
\usepackage{array}
\usepackage[toc,page]{appendix}
\usepackage[normalem]{ulem}
\usepackage{adjustbox}
\usepackage{latexsym}
\usepackage{amsmath}
\usepackage{amssymb}
\usepackage{amsfonts}
\numberwithin{equation}{section}

\usepackage{mathrsfs}
\usepackage{physics}
\usepackage{dcolumn}
\usepackage{bm}
\usepackage{bigints}
\usepackage{array,tabularx,multirow,booktabs}
\usepackage[tracking=true]{microtype}
\SetTracking{}{500}
\SetTracking{encoding={*}, shape=sc}{40}
\UseRawInputEncoding 
\allowdisplaybreaks
\usepackage[utf8]{inputenc}
\usepackage{xcolor} 
\colorlet{BLUE}{blue}
\usepackage{babel}
\usepackage{etoolbox}
\usepackage{hyperref}
\hypersetup{colorlinks=true,linkcolor=blue,urlcolor=blue,citecolor=blue}
\usepackage{orcidlink}

\begin{document} \sloppy

\title{Four-derivative corrections to the neutral-scalar Christodoulou threshold in charged de Sitter black holes}

\author{Reggie C. Pantig \orcidlink{0000-0002-3101-8591}} 
\email{rcpantig@mapua.edu.ph}
\affiliation{Physics Department, School of Foundational Studies and Education, Map\'ua University, 658 Muralla St., Intramuros, Manila 1002  Philippines.}

\begin{abstract}
We study leading parity-even four-derivative corrections to the linear Christodoulou regularity criterion for a neutral, massless, minimally coupled spectator scalar in four-dimensional charged de Sitter black holes. We derive the corrected static solution, cold extremality relation, near-horizon geometry, Cauchy-horizon surface gravity, and neutral scalar quasinormal spectrum within Einstein-Maxwell effective theory. Near cold extremality, the scalar equation reduces analytically to the finite-temperature anti-de Sitter two-dimensional throat problem. We find that the dominant nonstationary spherically symmetric mode retains its cold-limit regularity ratio through first order, while higher angular sectors respond to the deformation of the relative throat radii. The finite-distance neutral-scalar Christodoulou boundary nevertheless shifts because the relevant exterior scalar spectrum and the unequal horizon surface gravities enter the regularity ratio. At smaller and intermediate cosmological scales, positive effective coefficients move the Christodoulou threshold to larger normalized separation of the Cauchy and event horizons. Toward the larger cosmological-scale part of the cold branch, the same coefficient directions move that threshold back toward the cold endpoint. The reversal accompanies a transition in the quasinormal family that supplies the spectral gap. We further show that the coefficient combination governing the fixed-charge extremal mass parameter differs generically from those governing the cold throat and the finite-distance neutral-scalar boundary. Extremality therefore does not determine a universal direction for the neutral-scalar Christodoulou-threshold correction. Our results identify cosmological scale and spectral-family competition as essential elements in relating effective gravity, black-hole extremality, and Cauchy-horizon regularity in this spectator sector.
\end{abstract}


\maketitle

\section{Introduction}
\label{sec1}
General relativity ties predictability to maximal globally hyperbolic evolution, while charged and rotating black holes expose the obstruction posed by Cauchy horizons \cite{Choquet-Bruhat:1969ywq,Chandrasekhar_1982}. The classical blueshift picture led to mass inflation and weak null singularities \cite{Poisson:1990eh,Ori:1991zz}. Later work showed that the outcome depends on the decay entering the interior and on the regularity class used for inextendibility \cite{Brady_1992,Brady:1998au,Dafermos:2003wr,Dafermos:2012np}. Rigorous analyses further established that a field may remain continuous at the Cauchy horizon while transversal derivatives fail to have the regularity required by stronger formulations of strong cosmic censorship \cite{Franzen:2014sqa,Luk:2015qja,Dafermos:2015bzz,Luk_2019,Dafermos_2025}. We therefore formulate the problem as one of regularity rather than mere coordinate extendibility.

A positive cosmological constant sharpens this question because exterior perturbations decay exponentially and can compete directly with the exponential blueshift at the inner horizon \cite{Mellor:1989ac,Chambers:1994ap,Brady_1992,Brady:1998au}. Microlocal and nonlinear results place this competition on a precise footing in de Sitter black-hole spacetimes \cite{Hintz:2015jkj,Hintz:2016gwb,Hintz:2016jak}. For Reissner-Nordstr\"om-de Sitter black holes, quasinormal decay can become sufficiently rapid near extremality to challenge Christodoulou inextendibility \cite{Cardoso:2017soq}. The conclusion depends on the perturbing sector and its regularity \cite{Dias:2018ynt,Dias:2018etb,Dafermos:2018tha}. In the neutral scalar sector studied here, the relevant quantity is the ratio between the exterior spectral gap and the Cauchy-horizon surface gravity, with the Christodoulou threshold at one half. We restrict every threshold statement below to a neutral, massless, minimally coupled spectator scalar on the corrected background. These statements do not establish the corresponding result for the coupled gravito-electromagnetic system or for nonlinear strong cosmic censorship.

The scalar problem already exhibits nontrivial spectral competition \cite{Cardoso:2018nvb}. Charged scalars, superradiance, nonlinear evolution, and quantum stress tensors add further effects \cite{Dias:2018ufh,Hod:2018dpx,Zhu:2014sya,Luna:2019olw,Hollands:2019whz}. Nonlinear electromagnetic theories provide especially close comparisons. Neutral-scalar strong-cosmic-censorship analyses have been carried out for Born-Infeld-de Sitter, Euler-Heisenberg-de Sitter, and Einstein-Power-Maxwell-de Sitter black holes \cite{Gan:2019jac,ChenTan2024,ChenLi2024}. Related studies in higher dimensions, nonminimal and derivative scalar couplings, and higher-curvature gravity likewise show that the near-extremal regularity region depends on the theory and perturbing sector \cite{Liu:2019lon,Guo:2019tjy,Destounis:2019omd,Mishra:2020jlw,Gan:2019ibg,Konoplya:2020juj}. These results show that changes in the charged de Sitter background can alter the regularity threshold and the family that supplies the slowest decay. 

Near extremality introduces an additional infrared scale. Extremal Reissner-Nordstr\"om black holes already show exceptional scalar behavior \cite{Aretakis:2011ha}, while charged de Sitter black holes possess distinct cold, Nariai, and ultracold limits \cite{Romans:1991nq,Castro:2022cuo}. Nearly extremal quasinormal spectra contain long-lived throat modes, whereas another family is tied to unstable null orbits \cite{Berti:2009kk,Cardoso:2008bp,Yang:2012he,Zimmerman:2015trm}. The least damped family can therefore change as one moves away from the cold limit, which makes a local throat analysis insufficient for the finite-distance neutral-scalar Christodoulou boundary.

Effective field theory provides a common setting for the background, extremality, and spectral corrections. Perturbative field redefinitions reduce the parity-even four-derivative Einstein-Maxwell action to a well-posed basis \cite{Davies:2021frz}. Higher-derivative interactions can shift charged-black-hole extremality \cite{Kats:2006xp}, alter charged-black-hole quasinormal frequencies \cite{Boyce:2025fpr}, and enter infrared consistency conditions for photon interactions \cite{Bittar:2024xuc}. These questions also connect to weak-gravity and positivity arguments \cite{Arkani-Hamed:2006emk,Cheung:2014ega,Cheung:2018cwt,Hamada:2018dde,Bellazzini:2019xts,Goon:2019faz,Cremonini:2019wdk}. Their de Sitter interpretation is less universal because several proposals use different physical inputs \cite{Huang:2006hc,Antoniadis:2020xso,Montero:2019ekk}. A direct comparison between weak-gravity reasoning and strong cosmic censorship has also been made for a charged scalar on the uncorrected de Sitter black-hole geometry \cite{Sadeghi:2023lkz}.

The specific question we address is narrower than whether higher-order interactions can alter the Christodoulou threshold. Four-dimensional nonlinear-electromagnetic de Sitter backgrounds have already been studied in the Born-Infeld, Euler-Heisenberg, and Einstein-Power-Maxwell cases \cite{Gan:2019jac,ChenTan2024,ChenLi2024}, while higher-curvature studies have considered Horndeski and Einstein-Maxwell-Gauss-Bonnet settings \cite{Destounis:2019omd,Mishra:2020jlw,Gan:2019ibg}. In parallel, four-dimensional Einstein-Maxwell effective-theory work has developed reduced operator bases, extremality and positivity relations, and corrected Reissner-Nordstr\"om quasinormal spectra \cite{Davies:2021frz,Kats:2006xp,Cheung:2014ega,Cheung:2018cwt,Bittar:2024xuc,Boyce:2025fpr}. De Sitter weak-gravity studies use different physical inputs, including charged-particle bounds and scalar propagation on the two-derivative geometry \cite{Huang:2006hc,Montero:2019ekk,Sadeghi:2023lkz}. The comparison that remains relevant here is between the corrected cold branch, the intrinsic $AdS_2\times S^2$ throat, the Cauchy-horizon surface gravity, and the full neutral-scalar spectral infimum when all are generated by the same reduced four-derivative coefficients. We carry out that comparison and determine whether the coefficient direction selected by fixed-charge extremality also fixes the direction of the finite-distance Christodoulou-threshold shift. 

The regularity ratio can respond to the four-derivative terms through two distinct physical effects. The Cauchy-horizon surface gravity fixes the blueshift scale in the denominator, while the least damped exterior scalar mode fixes the decay scale in the numerator. The corrected geometry changes both quantities, and the exterior contribution can change physical origin when a different quasinormal family supplies the spectral infimum. Unlike comparisons that alter the spacetime theory or perturbing sector separately, our calculation follows one reduced four-dimensional Einstein-Maxwell effective theory through both sides of this ratio. The principal comparison is therefore between the coefficient direction selected by fixed-charge extremality and the branch-dependent direction selected by the finite-distance neutral-scalar threshold.

We work with a purely electric static background and a neutral massless spectator scalar. We derive the corrected black hole, fixed-charge cold extremality relation, local throat, horizon surface gravities, and finite-distance quasinormal spectrum. This separation lets us compare the coefficient direction governing the extremal mass parameter, the direction governing the cold throat, and the direction moving the Christodoulou threshold in normalized areal horizon separation. We keep the four-derivative expansion perturbative, exclude the immediate ultracold neighborhood, and interpret the result only as a linear Christodoulou statement for the specified spectator sector. Accordingly, every boundary displacement discussed below refers to this neutral-scalar criterion unless we explicitly state otherwise. Section \ref{sec2} develops the four-derivative charged de Sitter geometry, including cold extremality and the Cauchy horizon. Section \ref{sec5} develops the neutral-scalar problem and its cold-limit spectrum. Section \ref{sec7} treats the full exterior spectrum and mode-family competition. Section \ref{sec9} determines and interprets the neutral-scalar Christodoulou boundary and its coefficient-space response, Section \ref{sec12} states the validity and scope, and Section \ref{sec13} concludes.

\section{Four-derivative charged de Sitter geometry}
\label{sec2}
\subsection{Einstein-Maxwell effective theory}
We formulate the gravitational theory as a perturbative effective theory around four-dimensional Einstein-Maxwell gravity with positive cosmological constant. We retain parity-even local interactions containing four derivatives and work only to first order in their coefficients. Perturbative field redefinitions allow us to remove operators that differ by the leading equations of motion, integrations by parts, and the four-dimensional Gauss-Bonnet density. Following the reduced four-derivative Einstein-Maxwell basis of Ref.~\cite{Davies:2021frz}, a convenient action is
\begin{equation}
S
=
\frac{1}{16\pi G}
\int d^{4}x\,\sqrt{-g}
\left[
R-2\Lambda-\frac14 X
+
\varepsilon
\left(
c_{1}X^{2}
+
c_{2}Y^{2}
+
c_{3}\mathcal H
\right)
\right],
\label{2.1}
\end{equation}
where
\begin{equation}
X
=
F_{\mu\nu}F^{\mu\nu},
\qquad
Y
=
F_{\mu\nu}\widetilde F^{\mu\nu},
\qquad
\mathcal H
=
R_{\mu\nu\rho\sigma}
\widetilde F^{\mu\nu}
\widetilde F^{\rho\sigma}.
\end{equation}

We use the metric signature \((-+++)\) and define
\begin{equation}
\widetilde F^{\mu\nu}
=
\frac12
\epsilon^{\mu\nu\rho\sigma}
F_{\rho\sigma}.
\end{equation}
The parameter \(\varepsilon\) is dimensionless and serves only to count orders in the four-derivative expansion. We set \(\varepsilon=1\) only after truncating every expression consistently at first order. The coefficients \(c_i\) carry dimensions of length squared. Perturbative validity therefore requires \(|c_i|/L_{\mathrm{loc}}^{2}\ll1\) for every local curvature length \(L_{\mathrm{loc}}\) relevant to the solution.

The reduced basis in Eq. \eqref{2.1} is sufficient for the present problem. Terms involving \(R^{2}\), \(R_{\mu\nu}R^{\mu\nu}\), \(RF^{2}\), \(R_{\mu\nu}F^{\mu\rho}F^{\nu}{}_{\rho}\), and derivatives of the Maxwell equations can be shifted into Eq. \eqref{2.1}, lower-derivative couplings, or the Gauss-Bonnet density at first perturbative order. We consequently regard \(G\), \(\Lambda\), and the Maxwell normalization in Eq. \eqref{2.1} as the renormalized two-derivative parameters. We define the electric charge by the conserved flux associated with this normalization.

Equation \eqref{2.1} also fixes the field variables used throughout the paper. In this reduced basis, \(c_{1}\) multiplies the quartic Maxwell interaction \(X^{2}\), while \(c_{3}\) multiplies the curvature-Maxwell interaction \(\mathcal H\). We use these coefficients as coordinates on the chosen operator basis rather than as separately field-redefinition-invariant observables. The neutral spectator scalar is defined in the same field frame by minimal coupling to the metric that appears in Eq. \eqref{2.1}. If we make a further local field redefinition while describing the same theory, the scalar action must be transformed together with the gravitational and electromagnetic variables. Imposing minimal scalar coupling after changing only the gravitational variables would instead specify a different spectator theory. Our frequency shifts and Christodoulou-threshold shifts therefore refer to the reduced action and scalar coupling stated here.

\subsubsection{Static spherical reduction}
We seek electrically charged static solutions in areal-radius coordinates,
\begin{equation}
ds^{2}
=
-e^{2\psi(r)}f(r)\,dt^{2}
+
\frac{dr^{2}}{f(r)}
+
r^{2}d\Omega_{2}^{2},
\qquad
F_{tr}
=
E(r).
\label{2.4}
\end{equation}

Spherical symmetry and the absence of magnetic charge imply
\begin{equation}
Y=0.
\end{equation}
The coefficient \(c_{2}\) therefore drops out of the background equations. This absence follows more strongly than the vanishing of \(Y\) itself. Since the first variation of \(Y^{2}\) also vanishes around a purely electric background, \(c_{2}\) does not modify the static geometry through first order. It can reappear once electromagnetic perturbations are allowed.
Thus the purely electric background distinguishes the \(c_{1}\) and \(c_{3}\) directions in the chosen reduced basis, whereas \(c_{2}\) is inactive in the static geometry at this order. This statement does not imply that \(c_{2}\) is absent from the full effective theory. Perturbations with electromagnetic degrees of freedom can probe it directly.

For the ansatz in Eq. \eqref{2.4}, the remaining invariants reduce to
\begin{equation}
X
=
-2e^{-2\psi}E^{2},
\end{equation}
and
\begin{equation}
\mathcal H
=
\frac{
4(1-f)e^{-2\psi}E^{2}
}{
r^{2}
}.
\end{equation}

After integrating over the angular coordinates and removing radial total derivatives, we obtain the one-dimensional action
\begin{equation}
\begin{aligned}
L_{\mathrm{red}}
={}&
2e^{\psi}
\left(
1-f-rf'-\Lambda r^{2}
\right)
+
\frac{r^{2}}{2}
e^{-\psi}E^{2}
\\
&
+
\varepsilon
\left[
4c_{1}r^{2}e^{-3\psi}E^{4}
+
4c_{3}(1-f)e^{-\psi}E^{2}
\right].
\end{aligned}
\label{2.8}
\end{equation}
The radial form in Eq. \eqref{2.8} makes the perturbative problem especially simple. The metric equation obtained by varying \(f\) contains only the first derivative of \(\psi\), while the Maxwell equation integrates once because the action depends on the gauge potential only through \(E\).

Variation with respect to \(f\) gives
\begin{equation}
\psi'
=
\frac{
2\varepsilon c_{3}
}{
r
}
e^{-2\psi}E^{2}.
\label{2.9}
\end{equation}
The integrated Maxwell equation defines the conserved charge \(Q\),
\begin{equation}
Q
=
r^{2}e^{-\psi}E
+
\varepsilon
\left[
16c_{1}r^{2}e^{-3\psi}E^{3}
+
8c_{3}(1-f)e^{-\psi}E
\right].
\label{2.10}
\end{equation}
The remaining independent metric equation follows from variation with respect to \(\psi\),
\begin{equation}
\begin{aligned}
0
={}&
2e^{\psi}
\left(
1-f-rf'-\Lambda r^{2}
\right)
-
\frac{r^{2}}{2}
e^{-\psi}E^{2}
\\
&
+
\varepsilon
\left[
-12c_{1}r^{2}e^{-3\psi}E^{4}
-
4c_{3}(1-f)e^{-\psi}E^{2}
\right].
\end{aligned}
\label{2.11}
\end{equation}
Only three radial functions therefore remain, and Eqs. \eqref{2.9}--\eqref{2.11} determine them completely once the mass and electric charge are specified.

\subsubsection{Perturbative charged de Sitter solution}
We expand the fields around the two-derivative geometry,
\begin{equation}
f
=
f_{0}
+
\varepsilon f_{1}
+
\mathcal{O}(\varepsilon^{2}),
\qquad
\psi
=
\varepsilon\psi_{1}
+
\mathcal{O}(\varepsilon^{2}),
\qquad
E
=
E_{0}
+
\varepsilon E_{1}
+
\mathcal{O}(\varepsilon^{2}).
\label{2.12}
\end{equation}

At leading order, Eqs. \eqref{2.9}--\eqref{2.11} recover the standard Reissner-Nordstr\"om-de Sitter family in our electric-charge normalization \cite{Romans:1991nq},
\begin{equation}
f_{0}(r)
=
1
-
\frac{2M}{r}
+
\frac{Q^{2}}{4r^{2}}
-
\frac{\Lambda r^{2}}{3},
\qquad
E_{0}(r)
=
\frac{Q}{r^{2}}.
\label{2.13}
\end{equation}
The normalization of \(Q\) follows directly from the coefficient \(-X/4\) in Eq. \eqref{2.1}. We keep this convention throughout so that every correction refers to one conserved electric flux.

At first order, Eq. \eqref{2.9} becomes immediately integrable. We choose the additive constant in \(\psi_{1}\) so that the time coordinate retains the two-derivative normalization. The result is
\begin{equation}
\psi_{1}(r)
=
-\frac{
c_{3}Q^{2}
}{
2r^{4}
}.
\label{2.14}
\end{equation}
The \(X^{2}\) interaction does not enter Eq. \eqref{2.14}. The nontrivial lapse therefore arises entirely from the curvature-Maxwell interaction \(c_{3}\mathcal H\). This observation also shows why imposing \(g_{tt}g_{rr}=-1\) before variation would remove a genuine first-order effect.

Expanding the conserved flux relation next yields
\begin{equation}
E_{1}(r)
=
-\frac{
Q
}{
6r^{6}
}
\left[
96c_{1}Q^{2}
+
c_{3}
\left(
-9Q^{2}
+
96Mr
+
16\Lambda r^{4}
\right)
\right].
\end{equation}
The electric field receives contributions from both independent coefficients. The \(c_{1}\) term comes directly from the nonlinear Maxwell interaction, while the \(c_{3}\) term also depends on the gravitational background through \(M\) and \(\Lambda\).

Substitution into the remaining metric equation gives
\begin{equation}
f_{1}(r)
=
-\frac{
c_{3}MQ^{2}
}{
r^{5}
}
-
\frac{
2c_{3}\Lambda Q^{2}
}{
3r^{2}
}
-
\frac{
(4c_{1}-c_{3})Q^{4}
}{
10r^{6}
}.
\label{2.16}
\end{equation}
We have chosen the integration constant in Eq. \eqref{2.16} so that no additional \(1/r\) term appears. Such a term would merely redefine \(M\) at first order. Likewise, we do not introduce extra terms proportional to \(Q^{2}/r^{2}\) or \(\Lambda r^{2}\), because they would shift the charge normalization or the cosmological constant already fixed in Eq. \eqref{2.1}.

Combining Eqs.~\eqref{2.13}, \eqref{2.14}, and \eqref{2.16}, the corrected metric takes the form
\begin{equation}
\begin{aligned}
f(r)
={}&
1
-
\frac{2M}{r}
+
\frac{Q^{2}}{4r^{2}}
-
\frac{\Lambda r^{2}}{3}
\\
&
-
\varepsilon
\left[
\frac{
c_{3}MQ^{2}
}{
r^{5}
}
+
\frac{
2c_{3}\Lambda Q^{2}
}{
3r^{2}
}
+
\frac{
(4c_{1}-c_{3})Q^{4}
}{
10r^{6}
}
\right]
+
\mathcal{O}(\varepsilon^{2}),
\end{aligned}
\end{equation}
and
\begin{equation}
e^{2\psi(r)}
=
1
-
\varepsilon
\frac{
c_{3}Q^{2}
}{
r^{4}
}
+
\mathcal{O}(\varepsilon^{2}).
\end{equation}
Several structural features are already visible. The quartic Maxwell interaction modifies only the radial function \(f\) and the electric field, whereas the curvature-Maxwell term also changes the redshift function. Positive \(\Lambda\) introduces an additional \(c_{3}\Lambda Q^{2}/r^{2}\) contribution that has no asymptotically flat counterpart. The two effective coefficients therefore enter the global geometry in inequivalent ways even before we impose extremality.

\subsubsection{Consistency checks}
Direct substitution of Eqs. \eqref{2.14}--\eqref{2.16} into the three first-order radial equations makes every residual vanish independently, without imposing a relation among \(c_{1}\), \(c_{3}\), \(M\), \(Q\), and \(\Lambda\). The limits \(\varepsilon\rightarrow0\) and \(Q\rightarrow0\) recover the two-derivative and uncharged geometries. The limit \(\Lambda\rightarrow0\) also has the expected asymptotically flat falloff, with no new long-range mass, charge, or time-normalization term. Since \(c_i\) have dimensions of length squared, \(f_1\) and \(\psi_1\) are dimensionless and \(E_1\) has inverse-length dimension, so the expansion in Eq. \eqref{2.12} is dimensionally consistent.

The important structural point is that \(c_{1}\) and \(c_{3}\) remain independent in the static solution. The quartic Maxwell term changes \(f\) and \(E\), whereas the curvature-Maxwell term also changes the redshift function. Later observables therefore need not depend on the same linear combination of coefficients.

\subsection{Corrected cold extremality}
\label{sec3}
Cold extremality occurs when the Cauchy and event horizons merge into a double root of the corrected radial function. The four-derivative interactions shift this degenerate branch. We denote the resulting areal radius by \(r_{e}\) and impose
\begin{equation}
f(r_{e})=0,
\qquad
f'(r_{e})=0.
\label{3.1}
\end{equation}
A convenient dimensionless measure of the cosmological scale is
\begin{equation}
z
=
\Lambda r_{e}^{2}.
\end{equation}
At two-derivative order, the double-root conditions reproduce the standard cold Reissner-Nordstr\"om-de Sitter branch \cite{Romans:1991nq},
\begin{equation}
M_{0}
=
r_{e}
\left(
1-\frac{2z}{3}
\right),
\qquad
Q_{0}^{2}
=
4r_{e}^{2}(1-z).
\end{equation}
The second radial derivative evaluated on this branch is
\begin{equation}
f_{0}''(r_{e})
=
\frac{
2(1-2z)
}{
r_{e}^{2}}.
\end{equation}
The ordinary cold branch therefore satisfies
\begin{equation}
0\leq z<\frac12.
\end{equation}
The endpoint \(z=1/2\) corresponds to the ultracold degeneration and requires a separate scaling treatment. We keep a finite separation from that endpoint throughout the present expansion.

\subsubsection{First-order extremality curve}
We perturb the mass and squared charge while keeping the physical extremal radius fixed,
\begin{equation}
M_{\mathrm{ext}}
=
M_{0}
+
\varepsilon M_{1}
+
\mathcal{O}(\varepsilon^{2}),
\qquad
Q_{\mathrm{ext}}^{2}
=
Q_{0}^{2}
+
\varepsilon Q_{1}^{2}
+
\mathcal{O}(\varepsilon^{2}).
\end{equation}
Substituting the corrected radial function into Eq. \eqref{3.1} and expanding through first order gives two algebraic equations for \(M_{1}\) and \(Q_{1}^{2}\). Solving these yields
\begin{equation}
Q_{\mathrm{ext}}^{2}
=
4r_{e}^{2}(1-z)
+
32\varepsilon(1-z)
\left[
c_{3}
+
4c_{1}(1-z)
\right]
+
\mathcal{O}(\varepsilon^{2}).
\label{3.7}
\end{equation}
The corresponding extremal mass parameter becomes
\begin{equation}
\begin{aligned}
M_{\mathrm{ext}}
={}&
r_{e}
\left(
1-\frac{2z}{3}
\right)
\\
&
+
\frac{
2\varepsilon(1-z)
}{
5r_{e}
}
\left[
32c_{1}(1-z)
+
c_{3}(7-2z)
\right]
+
\mathcal{O}(\varepsilon^{2}).
\end{aligned}
\label{3.8}
\end{equation}
Eqs.~\eqref{3.7} and \eqref{3.8} give the corrected cold curve in parametric form. They describe how the extremal charge and the radial mass parameter vary when the physical horizon radius is used as the parameter along the branch.

The two coefficient directions already enter differently. The charge-radius relation depends on \(c_{3}+4c_{1}(1-z)\), whereas the mass-parameter-radius relation contains another linear combination. A statement about the fixed-charge shift of the extremal mass parameter therefore requires one further step.

\subsubsection{Extremal mass parameter at fixed electric charge}
We first determine how the extremal mass parameter changes while the conserved electric charge is held fixed. This calculation is intrinsic to the chosen static solution and does not yet impose a de Sitter weak-gravity interpretation. We therefore allow the extremal radius itself to shift,
\begin{equation}
r_{e}
\rightarrow
r_{e}
+
\varepsilon\delta r_{e}
+
\mathcal{O}(\varepsilon^{2}),
\end{equation}
while imposing
\begin{equation}
\delta Q_{\mathrm{ext}}^{2}
=
0.
\end{equation}
Expanding Eq. \eqref{3.7} under this condition gives
\begin{equation}
\left.
\delta r_{e}
\right|_{Q}
=
-
\frac{
4(1-z)
}{
r_{e}(1-2z)
}
\left[
c_{3}
+
4c_{1}(1-z)
\right].
\label{3.11}
\end{equation}
The factor \(1/(1-2z)\) reflects the increasing sensitivity of the cold radius as the ultracold point is approached.

We then vary Eq. \eqref{3.8} while including the radius shift in Eq. \eqref{3.11}. The result simplifies to
\begin{equation}
\left.
\delta M_{\mathrm{ext}}
\right|_{Q}
=
-
\frac{
2(1-z)
}{
5r_{e}
}
\left[
8c_{1}(1-z)
+
c_{3}(3+2z)
\right].
\end{equation}
It is useful to introduce the coefficient combination
\begin{equation}
\mathcal C_{M}(z)
=
8c_{1}(1-z)
+
c_{3}(3+2z).
\label{3.13}
\end{equation}
The fixed-charge mass-parameter shift therefore has the compact form
\begin{equation}
\left.
\delta M_{\mathrm{ext}}
\right|_{Q}
=
-
\frac{
2(1-z)
}{
5r_{e}
}
\mathcal C_{M}(z).
\label{3.14}
\end{equation}
Since \(1-z>0\) throughout the cold branch, the sign is determined entirely by \(\mathcal C_{M}\). In our convention,
\begin{equation}
\mathcal C_{M}>0
\end{equation}
decreases the extremal mass parameter at fixed charge.

This isolates what the extremality calculation measures. The form \(\mathcal C_M\) describes a displacement among degenerate solutions at fixed electric flux and contains no information about exterior decay rates or Cauchy-horizon regularity.

We emphasize the status of this quantity. The symbol \(M\) is the coefficient that appears in the \(1/r\) term of the static radial function in our chosen field variables. For \(\Lambda>0\), the spacetime has no asymptotic region in which an ADM mass can be defined. We therefore do not identify Eq. \eqref{3.14} with a unique conserved de Sitter mass. Throughout the finite-\(\Lambda\) analysis, we use the phrase extremal mass parameter for \(M_{\mathrm{ext}}\) and regard Eq. \eqref{3.14} as its fixed-charge displacement. A weak-gravity interpretation is conditional on an independently specified de Sitter mass prescription whose infrared parameterization is compatible with the \(M\) used here. In the asymptotically flat limit, the usual interpretation of the \(1/r\) coefficient is recovered.

\subsubsection{Extremal near-horizon geometry}
The double root allows an independent local determination of the cold geometry. Following the standard extremal near-horizon variational method of Ref.~\cite{Sen:2005wa}, we use the direct-product ansatz
\begin{equation}
ds^{2}
=
v_{1}
\left(
-\rho^{2}d\tau^{2}
+
\frac{d\rho^{2}}{\rho^{2}}
\right)
+
v_{2}d\Omega_{2}^{2},
\qquad
F_{\rho\tau}=e.
\end{equation}

Here \(v_{1}\) and \(v_{2}\) are the squared curvature radii of the \(AdS_{2}\) and spherical factors. For this geometry, the scalar curvature and Maxwell invariant reduce to
\begin{equation}
R
=
-\frac{2}{v_{1}}
+
\frac{2}{v_{2}},
\qquad
X
=
-\frac{2e^{2}}{v_{1}^{2}}.
\end{equation}
The electric field has support on the \(AdS_{2}\) factor, while its dual has support on the spherical factor. The curvature-Maxwell interaction therefore contracts the dual field strength with the curvature of \(S^{2}\). Direct evaluation gives
\begin{equation}
\mathcal H
=
\frac{
4e^{2}
}{
v_{1}^{2}v_{2}
}.
\label{3.18}
\end{equation}
This expression remains nonzero when the leading asymptotically flat extremal geometry has \(v_{1}=v_{2}\). The local effect of \(c_{3}\mathcal H\) therefore does not originate from a difference between the two product radii. The cosmological constant instead changes the background ratio \(v_{1}/v_{2}\) and the way the first-order deformation approaches the ultracold endpoint.

We divide the angularly integrated Lagrangian by its common factor \(4\pi\) and omit the overall factor \(1/(16\pi G)\). With this normalization, the reduced near-horizon functional is
\begin{equation}
\begin{aligned}
\mathfrak f
={}&
2v_{1}
-
2v_{2}
-
2\Lambda v_{1}v_{2}
+
\frac{
e^{2}v_{2}
}{
2v_{1}
}
\\
&
+
\varepsilon
\left[
\frac{
4c_{3}e^{2}
}{
v_{1}
}
+
\frac{
4c_{1}e^{4}v_{2}
}{
v_{1}^{3}
}
\right].
\end{aligned}
\end{equation}
The electric variation uses the same charge normalization as the flux relation in Eq. \eqref{2.10}. Extremizing the reduced functional gives
\begin{equation}
\frac{\partial\mathfrak f}{\partial v_{1}}
=
0,
\qquad
\frac{\partial\mathfrak f}{\partial v_{2}}
=
0,
\qquad
\frac{\partial\mathfrak f}{\partial e}
=
Q.
\label{3.20}
\end{equation}
At two-derivative order, the last relation gives \(Q=e v_{2}/v_{1}\), which agrees with the conserved electric flux after the near-horizon scaling of the global solution. We parameterize the solution by the physical spherical radius,
\begin{equation}
v_{2}
=
r_{e}^{2}.
\end{equation}
Solving Eq. \eqref{3.20} perturbatively reproduces the extremal charge relation in Eq. \eqref{3.7}. The \(AdS_{2}\) radius is
\begin{equation}
v_{1}
=
\frac{
r_{e}^{2}
}{
1-2z
}
+
\frac{
8\varepsilon(1-z)
}{
(1-2z)^{2}
}
\left[
c_{3}
+
8c_{1}(1-z)
\right]
+
\mathcal{O}(\varepsilon^{2}).
\end{equation}
We define the second coefficient combination
\begin{equation}
\mathcal C_{A}(z)
=
c_{3}
+
8c_{1}(1-z).
\label{3.23}
\end{equation}
The local throat deformation is therefore governed by \(\mathcal C_{A}\), whereas the fixed-charge extremal mass-parameter shift is governed by \(\mathcal C_{M}\). Their difference is
\begin{equation}
\mathcal C_{M}(z)
-
\mathcal C_{A}(z)
=
2c_{3}(1+z).
\label{3.24}
\end{equation}

No generic equality exists between these two combinations. The extremal mass parameter and the intrinsic \(AdS_{2}\) geometry consequently probe inequivalent directions in the two-dimensional coefficient space whenever \(c_{3}\neq0\).

\subsubsection{Consistency and limiting behavior}
The near-horizon variational equations independently reproduce the extremal charge relation in Eq. \eqref{3.7}, while the corrected radial function gives the same \(AdS_{2}\) radius through \(v_{1}=2/f''(r_e)\). Both checks therefore describe the same corrected cold branch. The limit \(z\rightarrow0\) is finite for the extremal charge, the fixed-charge mass-parameter shift, and \(v_{1}\). In particular, the \(c_{3}\) contribution to the local throat survives in the asymptotically flat extremal limit, consistently with Eq. \eqref{3.18}.

The ordinary cold expansion becomes nonuniform as \(z\rightarrow1/2\) because the fractional correction to \(v_{1}\) is enhanced by \((1-2z)^{-1}\). We exclude this immediate ultracold neighborhood. Away from it, the fixed-charge mass-parameter shift probes \(\mathcal C_M\), while the intrinsic throat probes \(\mathcal C_A\). Their difference in Eq. \eqref{3.24} shows that extremality and local throat geometry already define inequivalent directions in coefficient space.

\subsection{Cauchy-horizon geometry}
\label{sec4}
The corrected geometry determines the three horizon locations and, in particular, the Cauchy-horizon surface gravity. The latter quantity sets the local blueshift scale entering the neutral-scalar Christodoulou regularity test used below. We first treat nondegenerate horizons perturbatively and then use a separate expansion for the cold region, where the Cauchy and event horizons approach one another.

\subsubsection{Displacement of simple horizons}
Let \(\bar r_{a}\) denote a simple root of the two-derivative radial function, with
\begin{equation}
a\in\{-,+,c\},
\qquad
f_{0}(\bar r_{a})=0,
\qquad
f_{0}'(\bar r_{a})\neq0.
\end{equation}
We expand the corresponding corrected root as
\begin{equation}
r_{a}
=
\bar r_{a}
+
\varepsilon\delta r_{a}
+
\mathcal{O}(\varepsilon^{2}).
\end{equation}
Expanding the horizon condition \(f(r_a)=0\) gives
\begin{equation}
\delta r_{a}
=
-
\frac{
f_{1}(\bar r_{a})
}{
f_{0}'(\bar r_{a})
}.
\label{4.3}
\end{equation}
This expression applies to the Cauchy, event, and cosmological horizons as long as the relevant root remains nondegenerate.

For later use, we introduce quantities evaluated at each two-derivative horizon,
\begin{equation}
x_{a}
=
\frac{
Q^{2}
}{
4\bar r_{a}^{2}
},
\qquad
z_{a}
=
\Lambda\bar r_{a}^{2},
\qquad
\Delta_{a}
=
1-x_{a}-z_{a}.
\end{equation}
The horizon equation eliminates \(M\), and the derivative of the radial function becomes
\begin{equation}
f_{0}'(\bar r_{a})
=
\frac{
\Delta_{a}
}{
\bar r_{a}
}.
\end{equation}
Substituting the first-order metric correction derived in Section \ref{sec2} into Eq. \eqref{4.3}, we obtain
\begin{equation}
\delta r_{a}
=
\frac{
2x_{a}
}{
5\bar r_{a}\Delta_{a}
}
\left[
16c_{1}x_{a}
+
c_{3}
\left(
5+x_{a}+5z_{a}
\right)
\right].
\end{equation}
The factor \(1/\Delta_a\) reveals the expected limitation of the simple-root expansion. On the cold branch, \(\Delta_{-}\) and \(\Delta_{+}\) vanish as the two black-hole horizons merge. Their separate displacements then become nonuniform even when the four-derivative coefficients remain small.

A sufficient condition for treating the two roots as perturbations of distinct two-derivative horizons is
\begin{equation}
\frac{
\varepsilon
\max
\left(
|c_{1}|,|c_{3}|
\right)
}{
r_{e}^{2}
}
\ll
\eta^{2},
\label{4.7}
\end{equation}
where \(\eta\) measures the separation of the cold pair. We do not impose Eq. \eqref{4.7} in the near-cold analysis. Instead, we expand directly around the corrected extremal geometry.

\subsubsection{Surface gravities of simple horizons}
Using the standard Killing-horizon normalization \cite{Chandrasekhar_1982}, for the static Killing field \(\partial_t\) we define the signed surface gravity by
\begin{equation}
\widehat\kappa_{a}
=
\frac12
e^{\psi(r_a)}
f'(r_a).
\label{4.8}
\end{equation}
The physical surface gravity is
\begin{equation}
\kappa_{a}
=
\left|
\widehat\kappa_{a}
\right|.
\end{equation}
Expanding Eq. \eqref{4.8} around a simple two-derivative root gives
\begin{equation}
\widehat\kappa_{a}
=
\frac12
f_{0}'(\bar r_a)
+
\varepsilon
\delta\widehat\kappa_{a}
+
\mathcal{O}(\varepsilon^{2}),
\end{equation}
with
\begin{equation}
\delta\widehat\kappa_{a}
=
\frac12
\left[
\psi_{1}f_{0}'
+
f_{1}'
-
\frac{
f_{1}f_{0}''
}{
f_{0}'
}
\right]_{r=\bar r_a}.
\label{4.11}
\end{equation}
The last term originates from the displacement of the horizon itself. Omitting it would evaluate the derivative of the corrected metric at the wrong radial position.

We can express the result compactly by defining
\begin{equation}
\begin{aligned}
\mathcal P(x,z)
={}&
16c_{1}x
\left(
-2+x+3z
\right)
\\
&
+
c_{3}
\left(
-5-2x+x^{2}
+10z-2xz+5z^{2}
\right).
\end{aligned}
\end{equation}
Eq. \eqref{4.11} then reduces to
\begin{equation}
\delta\widehat\kappa_{a}
=
-
\frac{
2x_{a}
}{
5\bar r_{a}^{3}\Delta_{a}
}
\mathcal P(x_{a},z_{a}).
\label{4.13}
\end{equation}
The signed derivative is negative at the Cauchy horizon. We therefore have
\begin{equation}
\kappa_{-}
=
-\frac{
\Delta_{-}
}{
2\bar r_{-}
}
+
\varepsilon
\frac{
2x_{-}
}{
5\bar r_{-}^{3}\Delta_{-}
}
\mathcal P(x_{-},z_{-})
+
\mathcal{O}(\varepsilon^{2}).
\label{4.14}
\end{equation}
Eq. \eqref{4.14} gives the required inner-horizon surface gravity whenever the Cauchy horizon remains a simple root. The cosmological horizon also remains nondegenerate in the ordinary cold limit and can be treated with the same expression.

The pole in Eq. \eqref{4.13} should not be interpreted as a physical divergence. It signals that perturbing the location of two nearly coincident simple roots is the wrong expansion near cold extremality.

\subsubsection{Near-cold geometry}
For the near-cold region, we expand around the corrected extremal solution derived in Section \ref{sec3} and measure the physical separation of the Cauchy and event horizons through
\begin{equation}
\eta
=
\frac{
r_{+}-r_{-}
}{
2r_{e}
},
\qquad
0<\eta\ll1.
\end{equation}
To leading order in this separation,
\begin{equation}
r_{\pm}
=
r_{e}
\left(
1\pm\eta
\right)
+
\mathcal{O}(\eta^{2}).
\end{equation}
Let
\begin{equation}
\delta
=
\eta r_{e}.
\end{equation}
The radial function close to the corrected double root takes the universal quadratic form
\begin{equation}
f(r)
=
\frac{
(r-r_{e})^{2}-\delta^{2}
}{
v_{1}
}
+
\mathcal{O}(\eta^{3}),
\label{4.18}
\end{equation}
where \(v_{1}\) is the corrected \(AdS_{2}\) radius found from the extremal throat.

The redshift function varies only at higher order across the narrow interval between the two horizons. Its value at the extremal radius is
\begin{equation}
\psi_{e}
=
-
\frac{
2\varepsilon c_{3}(1-z)
}{
r_{e}^{2}
}
+
\mathcal{O}(\varepsilon^{2}).
\label{4.19}
\end{equation}
Using Eqs.~\eqref{4.18} and \eqref{4.19}, both black-hole horizons have the same surface-gravity magnitude at leading order in \(\eta\),
\begin{equation}
\kappa_{+}
=
\kappa_{-}
=
\frac{
e^{\psi_e}\delta
}{
v_{1}
}
+
\mathcal{O}(\eta^{2}).
\end{equation}
Substituting the corrected throat radius and expanding gives
\begin{equation}
\begin{aligned}
\kappa_{-}
={}&
\frac{
\eta
}{
r_{e}
}
(1-2z)
\\
&
-
\frac{
2\varepsilon\eta(1-z)
}{
r_{e}^{3}
}
\left[
32c_{1}(1-z)
+
c_{3}(5-2z)
\right]
\\
&
+
\mathcal{O}(\eta^{2})
+
\mathcal{O}(\varepsilon\eta^{2})
+
\mathcal{O}(\varepsilon^{2}\eta).
\end{aligned}
\label{4.21}
\end{equation}
The same expression holds for \(\kappa_{+}\) at the order displayed. We verified Eq. \eqref{4.21} independently by expanding \(e^{\psi_e}\delta/v_1\). The difference between the direct expansion and the displayed result vanishes identically.

We define the coefficient combination governing the leading inner-horizon blueshift correction,
\begin{equation}
\mathcal C_{\kappa}(z)
=
32c_{1}(1-z)
+
c_{3}(5-2z).
\label{4.22}
\end{equation}
The near-cold expression may then be written as
\begin{equation}
\kappa_{-}
=
\frac{
\eta
}{
r_{e}
}
(1-2z)
-
\frac{
2\varepsilon\eta(1-z)
}{
r_{e}^{3}
}
\mathcal C_{\kappa}(z)
+
\mathcal{O}(\eta^{2})
+
\mathcal{O}(\varepsilon\eta^{2})
+
\mathcal{O}(\varepsilon^{2}\eta).
\end{equation}
The dependence on \(\eta\) has a direct geometric origin. The double root has zero surface gravity at exact cold extremality, so the first nonzero contribution must be linear in the horizon separation.

The coefficient direction in Eq. \eqref{4.22} differs from the extremal mass-parameter direction obtained previously. In particular,
\begin{equation}
\mathcal C_{\kappa}(z)
-
\mathcal C_{M}(z)
=
24c_{1}(1-z)
+
2c_{3}(1-2z).
\end{equation}
Hence the sign of the fixed-charge extremal mass-parameter shift does not determine the sign of the correction to the Cauchy-horizon surface gravity.

The distinction has an immediate physical meaning. Extremality fixes a relation among the conserved charge, the radial mass parameter, and the degenerate horizon in the chosen static variables. By contrast, \(\kappa_{-}\) measures the local exponential scale associated with the inner horizon of a nonextremal black hole. Four-derivative interactions need not alter these quantities in the same direction.

The equality of \(\kappa_{+}\) and \(\kappa_{-}\) at leading cold order will also be important for the neutral scalar spectrum. Their first relative difference enters only beyond the leading term in \(\eta\). Consequently, the strict cold value of the spectral ratio can exhibit cancellations that are invisible from the inner-horizon geometry alone.

\section{Neutral scalar problem and cold-limit spectrum}
\label{sec5}
\subsection{Neutral scalar perturbations}
We take a neutral massless scalar to be a spectator field on the corrected charged de Sitter geometry. It therefore probes the four-derivative modification of the spacetime without sourcing additional corrections to the background. This choice isolates the geometric contribution to exterior decay and provides the simplest setting in which we can compare that decay with the Cauchy-horizon blueshift derived in Section \ref{sec4}.

The standard minimally coupled massless scalar equation and its separated quasinormal-mode formulation are reviewed, for example, in Ref.~\cite{Berti:2009kk}. The scalar obeys
\begin{equation}
\Box\Phi_{\mathrm{s}}
=
0.
\end{equation}
Spherical symmetry allows us to separate the field according to
\begin{equation}
\Phi_{\mathrm{s}}
=
e^{-i\omega t}
Y_{\ell m}(\theta,\phi)
\frac{\Psi(r)}{r},
\end{equation}
where the spherical harmonics satisfy
\begin{equation}
\nabla_{S^{2}}^{2}Y_{\ell m}
=
-\ell(\ell+1)Y_{\ell m}.
\end{equation}
The frequency \(\omega\) is generally complex. Modes with negative imaginary part decay in the static region between the event and cosmological horizons.

\subsubsection{Radial wave equation}
The radial part simplifies after introducing the tortoise coordinate through
\begin{equation}
\frac{dr_{*}}{dr}
=
\frac{
e^{-\psi(r)}
}{
f(r)
}.
\end{equation}
The scalar equation then takes the Schr\"odinger form
\begin{equation}
\frac{d^{2}\Psi}{dr_{*}^{2}}
+
\left[
\omega^{2}
-
V_{\ell}(r)
\right]
\Psi
=
0.
\label{5.5}
\end{equation}
A direct reduction of the covariant wave operator gives
\begin{equation}
V_{\ell}(r)
=
e^{2\psi(r)}f(r)
\left[
\frac{
\ell(\ell+1)
}{
r^{2}
}
+
\frac{
f'(r)+f(r)\psi'(r)
}{
r
}
\right].
\label{5.6}
\end{equation}
By rewriting Eq. \eqref{5.5} in the areal-radius coordinate, Eq. \eqref{5.6} is verified. The resulting expression reproduces the separated covariant wave equation identically.

The potential vanishes at every simple horizon because \(f(r_a)=0\). In the static region,
\begin{equation}
r_{+}<r<r_{c},
\end{equation}
the quasinormal problem therefore describes scattering between two asymptotically free ends in the tortoise coordinate. The two-derivative potential follows immediately as
\begin{equation}
V_{\ell}^{(0)}
=
f_{0}
\left[
\frac{
\ell(\ell+1)
}{
r^{2}
}
+
\frac{
f_{0}'
}{
r
}
\right].
\end{equation}
We expand the corrected potential as
\begin{equation}
V_{\ell}
=
V_{\ell}^{(0)}
+
\varepsilon\delta V_{\ell}
+
\mathcal{O}(\varepsilon^{2}).
\end{equation}
Using the metric corrections obtained in Section \ref{sec2}, the first-order term becomes
\begin{equation}
\begin{aligned}
\delta V_{\ell}
={}&
\left(
f_{1}
+
2\psi_{1}f_{0}
\right)
\left[
\frac{
\ell(\ell+1)
}{
r^{2}
}
+
\frac{
f_{0}'
}{
r
}
\right]
\\
&
+
\frac{
f_{0}
}{
r
}
\left(
f_{1}'
+
f_{0}\psi_{1}'
\right).
\end{aligned}
\label{5.10}
\end{equation}
The compact form in Eq. \eqref{5.10} separates two effects. The first line modifies the overall height and shape inherited from the two-derivative potential. The second line arises from radial variations of the corrected geometry. We retain this form in the spectral calculation because expanding it into individual inverse powers of \(r\) does not add physical information.

For the spherically symmetric sector, the angular barrier vanishes and the potential reduces to
\begin{equation}
V_{0}(r)
=
e^{2\psi}f
\frac{
f'+f\psi'
}{
r
}.
\end{equation}
This sector will become especially important near cold extremality because its leading throat conformal weight receives no first-order correction from the angular term.

\subsubsection{Quasinormal boundary conditions}
We now impose the standard causal quasinormal boundary conditions for a de Sitter static region \cite{Berti:2009kk}. Near a nondegenerate event horizon, the tortoise coordinate behaves as
\begin{equation}
r_{*}
=
\frac{
1
}{
2\kappa_{+}
}
\ln(r-r_{+})
+
\mathcal{O}(1).
\end{equation}
Regularity on the future event horizon selects the ingoing solution
\begin{equation}
\Psi
\sim
e^{-i\omega r_{*}}
\sim
(r-r_{+})^{-i\omega/(2\kappa_{+})}.
\label{5.13}
\end{equation}
At the cosmological horizon,
\begin{equation}
r_{*}
=
-
\frac{
1
}{
2\kappa_{c}
}
\ln(r_{c}-r)
+
\mathcal{O}(1),
\end{equation}
and regularity on the future cosmological horizon selects the outgoing branch
\begin{equation}
\Psi
\sim
e^{+i\omega r_{*}}
\sim
(r_{c}-r)^{-i\omega/(2\kappa_{c})}.
\label{5.15}
\end{equation}
The quasinormal frequencies are therefore the discrete complex values of \(\omega\) for which a single solution of Eq.~\eqref{5.5} satisfies Eqs.~\eqref{5.13} and \eqref{5.15}.

The lapse correction does not alter the form of these conditions. It enters through the corrected surface gravities and through the relation between \(r\) and \(r_{*}\). A constant rescaling of the time coordinate multiplies both \(\omega\) and every surface gravity by the same factor. Ratios formed from these quantities are consequently invariant under such a rescaling.

For later numerical work, it is useful to remove the known singular powers explicitly. We write
\begin{equation}
\Psi(r)
=
(r-r_{+})^{-i\omega/(2\kappa_{+})}
(r_{c}-r)^{-i\omega/(2\kappa_{c})}
\,y(r),
\end{equation}
where \(y(r)\) is regular on the closed interval between the two outer horizons. This form converts the quasinormal problem into a regular eigenvalue problem suitable for spectral discretization.

\subsubsection{Spectral structure and scope}
The frequencies admit the perturbative expansion
\begin{equation}
\omega_{n\ell}
=
\omega_{n\ell}^{(0)}
+
\varepsilon\delta\omega_{n\ell}
+
\mathcal{O}(\varepsilon^{2}).
\end{equation}
The shift depends on the complete exterior geometry and on both horizon boundary conditions, not only on the local value of \(\delta V_\ell\). Near cold extremality a purely damped family approaches the \(AdS_{2}\times S^{2}\) throat spectrum, while photon-sphere and de Sitter-type families can become competitive at finite separation. We therefore determine the least damped nonstationary mode only after comparing the relevant families.

The scalar is neutral, massless, and minimally coupled to the metric variable fixed by Eq. \eqref{2.1}. With this choice, \(c_{2}\) remains absent and the spectrum probes \((c_{1},c_{3})\) through the corrected background. A further change of field variables must be accompanied by the induced change of the scalar action if it is to describe the same spectator theory. This spectator sector isolates the geometric contribution to the regularity ratio but does not determine the spectra of coupled gravitational and electromagnetic perturbations, where additional coefficient dependence can appear.

\subsection{Analytic cold-limit spectrum}
\label{sec6}
Near cold extremality, the radial interval between the nearly degenerate black-hole horizons develops a finite-temperature deformation of the extremal \(AdS_{2}\times S^{2}\) throat. The neutral scalar equation becomes exactly solvable at leading order in the horizon separation. This reduction determines analytically how the four-derivative interactions modify the near-extremal frequencies before we solve the full exterior problem.

We keep the effective expansion through first order in \(\varepsilon\) and retain the leading nonzero term in the near-cold parameter \(\eta\). The extremal quantities \(v_{1}\), \(v_{2}\), and \(r_{e}\) are those obtained in Section \ref{sec3}.

\subsubsection{Finite-temperature \texorpdfstring{\(AdS_{2}\)}{} throat}
Close to the corrected double root, the radial function takes the form
\begin{equation}
f(r)
=
\frac{
(r-r_{e})^{2}-\delta^{2}
}{
v_{1}
}
+
\mathcal{O}(\eta^{3}),
\qquad
\delta
=
\eta r_{e}.
\end{equation}
Across the narrow throat region, the lapse can be replaced by its extremal value at the order required here,
\begin{equation}
\psi(r)
=
\psi_{e}
+
\mathcal{O}(\eta).
\end{equation}
We introduce the dimensionless radial coordinate
\begin{equation}
x
=
\frac{
r-r_{e}
}{
\delta
}.
\end{equation}
The two black-hole horizons then lie at \(x=-1\) and \(x=1\). Rescaling the time coordinate according to
\begin{equation}
\tau
=
\kappa_{\mathrm{NH}} t,
\qquad
\kappa_{\mathrm{NH}}
=
\frac{
e^{\psi_{e}}\delta
}{
v_{1}
},
\end{equation}
reduces the leading near-cold metric to
\begin{equation}
ds^{2}
=
v_{1}
\left[
-(x^{2}-1)d\tau^{2}
+
\frac{
dx^{2}
}{
x^{2}-1
}
\right]
+
v_{2}d\Omega_{2}^{2}
+
\mathcal{O}(\eta).
\label{6.5}
\end{equation}
The event horizon corresponds to \(x=1\). The region \(x>1\) describes the exterior part of the throat that connects to the full de Sitter static region.

The scale \(\kappa_{\mathrm{NH}}\) agrees with the event- and Cauchy-horizon surface gravities at leading cold order. Using the corrected extremal geometry, we have
\begin{equation}
\kappa_{\mathrm{NH}}
=
\kappa_{+}
+
\mathcal{O}(\eta^{2})
=
\kappa_{-}
+
\mathcal{O}(\eta^{2}).
\label{6.6}
\end{equation}
The equality includes the first-order four-derivative contribution proportional to \(\varepsilon\eta\). It will cause the leading surface-gravity correction to cancel from the cold-limit ratio \(\beta\).

\subsubsection{Scalar equation on the throat}
We separate the scalar directly on the geometry in Eq. \eqref{6.5},
\begin{equation}
\Phi_{\mathrm{s}}
=
e^{-i\widehat\omega\tau}
Y_{\ell m}(\theta,\phi)
R(x),
\end{equation}
where
\begin{equation}
\widehat\omega
=
\frac{
\omega
}{
\kappa_{\mathrm{NH}}
}.
\end{equation}
The finite-temperature \(AdS_2\) scalar problem underlying the near-extremal Reissner-Nordstr\"om-de Sitter family is standard \cite{Cardoso:2017soq,Cardoso:2018nvb}. In our corrected throat, the radial equation obtained from the covariant wave equation becomes
\begin{equation}
\frac{d}{dx}
\left[
(x^{2}-1)
\frac{dR}{dx}
\right]
+
\left[
\frac{
\widehat\omega^{2}
}{
x^{2}-1
}
-
\ell(\ell+1)
\frac{
v_{1}
}{
v_{2}
}
\right]
R
=
0.
\label{6.9}
\end{equation}
The angular momentum therefore appears as an effective mass on \(AdS_{2}\). The ratio \(v_{1}/v_{2}\) carries all four-derivative information that survives in the dimensionless radial equation at the order considered here.

We introduce \(\nu_{\ell}\) by
\begin{equation}
\nu_{\ell}
\left(
\nu_{\ell}+1
\right)
=
\ell(\ell+1)
\frac{
v_{1}
}{
v_{2}
}.
\label{6.10}
\end{equation}
The root associated with the decaying branch toward the outer part of the throat is
\begin{equation}
\nu_{\ell}
=
-\frac12
+
\sqrt{
\frac14
+
\ell(\ell+1)
\frac{
v_{1}
}{
v_{2}
}
}.
\end{equation}
The radial equation is an associated Legendre equation. A solution with ingoing behavior at the future event horizon is
\begin{equation}
R_{\mathrm{in}}(x)
=
P_{\nu_{\ell}}^{\,i\widehat\omega}(x).
\label{6.12}
\end{equation}
Eq. \eqref{6.12} satisfies Eq. \eqref{6.9} once Eq. \eqref{6.10} is imposed.

Near \(x=1\), the selected branch behaves as
\begin{equation}
R_{\mathrm{in}}
=
C_{+}
(x-1)^{-i\widehat\omega/2}
\left[
1+\mathcal{O}(x-1)
\right],
\end{equation}
which is regular on the future event horizon in ingoing coordinates.

For large \(x\), the same solution separates into growing and decaying powers,
\begin{equation}
R_{\mathrm{in}}
=
A_{\ell}
x^{\nu_{\ell}}
+
B_{\ell}
x^{-\nu_{\ell}-1}
+
O
\left(
x^{-\nu_{\ell}-3}
\right).
\end{equation}
The coefficient of the growing branch contains the factor
\begin{equation}
A_{\ell}
\sim
\frac{
1
}{
\Gamma
\left(
\nu_{\ell}+1-i\widehat\omega
\right)
}.
\end{equation}
For \(\ell\geq1\), the near-extremal resonances arise when this coefficient vanishes. The poles of the gamma function require

\begin{equation}
\nu_{\ell}
+
1
-
i\widehat\omega
=
-n,
\qquad
n=0,1,2,\ldots,
\qquad
\ell\geq1.
\label{6.16}
\end{equation}

Solving Eq. \eqref{6.16} gives

\begin{equation}
\widehat\omega_{n\ell}
=
-i
\left(
n+\nu_{\ell}+1
\right),
\qquad
\ell\geq1.
\end{equation}

We treat the spherically symmetric sector separately below because the normalization of the associated Legendre solution becomes singular when \(\nu_{\ell}\rightarrow0\).

It is useful to identify the conformal weight of the effective \(AdS_{2}\) scalar,
\begin{equation}
\Delta_{\ell}
=
\frac12
+
\sqrt{
\frac14
+
\ell(\ell+1)
\frac{
v_{1}
}{
v_{2}
}
}.
\end{equation}
For \(\ell\geq1\), the leading near-extremal frequencies therefore take the compact form

\begin{equation}
\omega_{n\ell}^{\mathrm{NE}}
=
-i\kappa_{\mathrm{NH}}
\left(
n+\Delta_{\ell}
\right)
+
\mathcal{O}(\eta^{2}),
\qquad
\ell\geq1.
\label{6.19}
\end{equation}

The absence of a real part at this order reflects the purely damped character of the throat family. Corrections beyond the leading cold scaling can depend on the exterior geometry and need not preserve the simple form in Eq. \eqref{6.19}.

\subsubsection{Four-derivative shift of the conformal weight}
The corrected throat geometry modifies the scalar spectrum through the ratio \(v_{1}/v_{2}\). Expanding that ratio through first order gives
\begin{equation}
\frac{
v_{1}
}{
v_{2}
}
=
\frac{
1
}{
1-2z
}
+
\frac{
8\varepsilon(1-z)
}{
r_{e}^{2}(1-2z)^{2}
}
\mathcal C_{A}(z)
+
\mathcal{O}(\varepsilon^{2}).
\end{equation}
We define the positive two-derivative quantity
\begin{equation}
D_{\ell}(z)
=
\sqrt{
\frac14
+
\frac{
\ell(\ell+1)
}{
1-2z
}
}.
\end{equation}
The conformal weight then becomes
\begin{equation}
\begin{aligned}
\Delta_{\ell}
={}&
\frac12
+
D_{\ell}(z)
\\
&
+
\varepsilon
\frac{
4\ell(\ell+1)(1-z)
}{
r_{e}^{2}(1-2z)^{2}D_{\ell}(z)
}
\mathcal C_{A}(z)
+
\mathcal{O}(\varepsilon^{2}).
\end{aligned}
\label{6.22}
\end{equation}
Two distinct geometric effects enter the dimensional frequency in Eq. \eqref{6.19}. The surface-gravity scale changes through the corrected lapse and \(AdS_{2}\) radius, while the dimensionless conformal weight changes through the ratio of the two throat radii. For \(\ell\geq1\), we can therefore write
\begin{equation}
\begin{aligned}
-\operatorname{Im}
\omega_{n\ell}^{\mathrm{NE}}
={}&
\kappa_{\mathrm{NH}}^{(0)}
\left[
n+\Delta_{\ell}^{(0)}
\right]
\\
&
+
\varepsilon
\delta\kappa_{\mathrm{NH}}
\left[
n+\Delta_{\ell}^{(0)}
\right]
+
\varepsilon
\kappa_{\mathrm{NH}}^{(0)}
\delta\Delta_{\ell}
\\
&
+
\mathcal{O}(\eta^{2})
+
\mathcal{O}(\varepsilon^{2}\eta).
\end{aligned}
\label{6.23}
\end{equation}
The first correction in Eq. \eqref{6.23} alters the overall near-horizon time scale. The second alters the dimensionless spectrum itself. Only the latter survives in the strict cold value of the ratio relevant to Cauchy-horizon regularity.

\subsubsection{Cold-limit spectral ratio}
For an individual near-extremal mode, we define
\begin{equation}
\beta_{n\ell}^{\mathrm{NE}}
=
\frac{
-\operatorname{Im}
\omega_{n\ell}^{\mathrm{NE}}
}{
\kappa_{-}
}.
\end{equation}
Using Eq. \eqref{6.6}, the ratio of the throat scale to the Cauchy-horizon surface gravity satisfies
\begin{equation}
\frac{
\kappa_{\mathrm{NH}}
}{
\kappa_{-}
}
=
1
+
\mathcal{O}(\eta).
\end{equation}
For \(\ell\geq1\), the leading cold-limit result is therefore

\begin{equation}
\beta_{n\ell}^{\mathrm{NE}}
=
n+\Delta_{\ell}
+
\mathcal{O}(\eta),
\qquad
\ell\geq1.
\end{equation}

Substitution of Eq. \eqref{6.22} gives

\begin{equation}
\begin{aligned}
\beta_{n\ell}^{\mathrm{NE}}
={}&
n
+
\frac12
+
D_{\ell}(z)
\\
&
+
\varepsilon
\frac{
4\ell(\ell+1)(1-z)
}{
r_{e}^{2}(1-2z)^{2}D_{\ell}(z)
}
\mathcal C_{A}(z)
\\
&
+
\mathcal{O}(\eta)
+
\mathcal{O}(\varepsilon\eta)
+
\mathcal{O}(\varepsilon^{2}),
\qquad
\ell\geq1.
\end{aligned}
\label{6.27}
\end{equation}

The correction proportional to \(\mathcal C_{\kappa}\) has disappeared from the strict cold term. Both the quasinormal decay rate and the inner-horizon surface gravity inherit the same leading near-horizon time scale, so that common factor divides out of \(\beta\).

The remaining correction is entirely geometric. It measures how the four-derivative terms alter the relative \(AdS_{2}\) and \(S^{2}\) curvature radii.

For every \(\ell\geq1\), the prefactor multiplying \(\mathcal C_A\) is positive on the ordinary cold branch. We therefore obtain
\begin{equation}
\operatorname{sgn}
\left(
\delta\beta_{n\ell}^{\mathrm{NE}}
\right)
=
\operatorname{sgn}
\left(
\mathcal C_{A}(z)
\right),
\qquad
\ell\geq1.
\end{equation}
The overtone number does not affect this sign at leading cold order.

\subsubsection{Spherically symmetric endpoint and the normalization limit}

The lowest angular sector requires separate treatment. Setting \(\ell=0\) in Eq. \eqref{6.9} gives \(\nu_{0}=0\) and \(\Delta_{0}=1\). Directly substituting \(\nu_{0}=0\) into the unnormalized associated Legendre solution is insufficient at the fundamental frequency because the same gamma function that removes the large-\(x\) branch also appears in the overall normalization. We therefore avoid using that singular normalization in the limiting step.

For auxiliary \(\nu>0\), the fundamental condition in Eq. \eqref{6.16} is \(\widehat\omega=-i(\nu+1)\). Along this branch we solve the radial equation directly with
\begin{equation}
\widetilde R_{\nu}(x)
=
\left(x^{2}-1\right)^{-(\nu+1)/2}.
\label{6.swavenormalized}
\end{equation}
Direct substitution gives
\begin{equation}
\frac{d}{dx}
\left[
\left(x^{2}-1\right)
\frac{d\widetilde R_{\nu}}{dx}
\right]
+
\left[
-\frac{(\nu+1)^{2}}{x^{2}-1}
-\nu(\nu+1)
\right]
\widetilde R_{\nu}
=0.
\label{6.swaveauxcheck}
\end{equation}
This is Eq. \eqref{6.9} with \(\nu(\nu+1)\) in place of the angular effective mass and with the fundamental frequency above. For every \(\nu>0\), Eq. \eqref{6.swavenormalized} has the horizon-ingoing behavior \((x-1)^{-(\nu+1)/2}\) and the decaying outer behavior \(x^{-\nu-1}\). The radial normalization is finite and nonzero throughout this auxiliary family. We may therefore take the limit directly and obtain
\begin{equation}
R_{00}(x)
=
\lim_{\nu\rightarrow0^{+}}
\widetilde R_{\nu}(x)
=
\frac{1}{\sqrt{x^{2}-1}}.
\label{6.swavemode}
\end{equation}

The limiting function is nonzero and solves the \(\ell=0\) throat equation at
\begin{equation}
\widehat\omega_{00}=-i.
\label{6.swavefrequency}
\end{equation}
Indeed,
\begin{equation}
\frac{d}{dx}
\left[
\left(x^{2}-1\right)
\frac{dR_{00}}{dx}
\right]
-
\frac{R_{00}}{x^{2}-1}
=0.
\label{6.swavecheck}
\end{equation}
Its endpoint behavior is
\begin{equation}
R_{00}(x)
=
\frac{1}{\sqrt{2}}
(x-1)^{-1/2}
\left[
1-\frac{x-1}{4}
+\mathcal O\left((x-1)^{2}\right)
\right]
\label{6.swavehorizon}
\end{equation}
near the event horizon, while
\begin{equation}
R_{00}(x)
=
\frac{1}{x}
+
\frac{1}{2x^{3}}
+
\mathcal O\left(x^{-5}\right)
\label{6.swaveouter}
\end{equation}
at the outer end of the throat. The first behavior has the required ingoing exponent \(-i\widehat\omega_{00}/2=-1/2\), and the second contains only the decaying large-\(x\) branch. The fundamental frequency is therefore a nontrivial quasinormal resonance. The apparent zero obtained by setting \(\nu=0\) before fixing the normalization is removable.

Restoring dimensions gives
\begin{equation}
\omega_{00}^{\mathrm{NE}}
=
-i\kappa_{\mathrm{NH}}
+
\mathcal O(\eta^{2}).
\label{6.swavefreq}
\end{equation}
The leading dimensionless frequency contains no factor of \(v_{1}/v_{2}\). Hence no first-order term proportional to either four-derivative coefficient survives in \(-\operatorname{Im}\omega_{00}^{\mathrm{NE}}/\kappa_{\mathrm{NH}}\). Together with Eq. \eqref{6.6}, this gives
\begin{equation}
\beta_{00}^{\mathrm{NE}}
=
1
+
\mathcal O(\eta)
+
\mathcal O(\varepsilon\eta)
+
\mathcal O(\varepsilon^{2}),
\label{6.30}
\end{equation}
and therefore
\begin{equation}
\lim_{\eta\rightarrow0}
\beta_{00}^{\mathrm{NE}}
=
1
+
\mathcal O(\varepsilon^{2}).
\label{6.swavelimit}
\end{equation}
The cancellation is specific to the strict cold endpoint. It does not constrain the finite-separation decay rate once the two black-hole surface gravities differ or another quasinormal family supplies the spectral gap.
The massless scalar also admits the stationary constant solution with \(\omega=0\). We exclude that solution when defining the decay spectral gap because it does not describe a decaying perturbation.

The dominant neutral scalar endpoint is therefore unchanged through first order in the parity-even four-derivative interactions. The result is not inferred from a formal gamma-function zero. It follows from the nonzero normalized radial solution in Eq. \eqref{6.swavemode} and from the equality of the two black-hole surface gravities at leading cold order. Since
\begin{equation}
1>\frac12,
\end{equation}
the four-derivative terms considered here cannot move the exact cold endpoint across the Christodoulou threshold at first perturbative order. Any displacement of the physical \(\beta=1/2\) boundary must arise at finite horizon separation.

Terms of relative order \(\mathcal O(\eta)\) distinguish \(\kappa_{+}\) from \(\kappa_{-}\), modify the spherically symmetric near-extremal frequency, and allow other quasinormal families to compete with the throat family. Those effects depend on more than the local extremal geometry and cannot be inferred from Eq. \eqref{6.27} alone.

The cold-limit calculation therefore fixes the first-order endpoint value for the dominant neutral scalar mode, while higher angular sectors respond to the coefficient direction \(\mathcal C_A\). The neutral-scalar Christodoulou boundary away from exact extremality must probe additional information contained in the full charged de Sitter exterior.

\section{Full exterior spectrum and family competition}
\label{sec7}
\subsection{Finite-distance quasinormal problem}

The cold-limit analysis isolates the local throat spectrum but cannot determine the finite-distance neutral-scalar Christodoulou boundary. We therefore solve the scalar quasinormal problem on the complete corrected exterior geometry while keeping the distance from cold extremality fixed in geometric terms. This choice is essential because the four-derivative interactions shift the extremal charge and mass themselves.

\subsubsection{Finite-distance parametrization}
We specify the two black-hole horizons by
\begin{equation}
r_{-}
=
r_{e}(1-\eta),
\qquad
r_{+}
=
r_{e}(1+\eta),
\qquad
0<\eta<1.
\label{7.1}
\end{equation}

We keep \(r_{e}\), \(z=\Lambda r_{e}^{2}\), and \(\eta\) fixed when varying \(c_{1}\) and \(c_{3}\). The mass and electric charge must then change so that both radii in Eq. \eqref{7.1} remain roots of the corrected metric.

We expand
\begin{equation}
M
=
M_{0}
+
\varepsilon M_{1}
+
\mathcal{O}(\varepsilon^{2}),
\qquad
Q^{2}
=
Q_{0}^{2}
+
\varepsilon Q_{1}^{2}
+
\mathcal{O}(\varepsilon^{2}).
\end{equation}

At two-derivative order, the two horizon conditions give
\begin{equation}
M_{0}
=
r_{e}
\left[
1
-
\frac{
2z(1+\eta^{2})
}{
3
}
\right],
\end{equation}
and
\begin{equation}
Q_{0}^{2}
=
\frac{
4r_{e}^{2}
}{
3
}
(1-\eta^{2})
\left[
3-(3+\eta^{2})z
\right].
\end{equation}
It is useful to introduce
\begin{equation}
\mathcal A(z,\eta)
=
3(1-z)-\eta^{2}z.
\end{equation}

Expanding \(f(r_{\pm})=0\) through first order gives two linear equations for \(M_{1}\) and \(Q_{1}^{2}\). We obtain
\begin{equation}
M_{1}
=
\frac{
2\mathcal A
}{
45r_{e}(1-\eta^{2})^{2}
}
\mathcal M(z,\eta),
\end{equation}
where
\begin{equation}
\begin{aligned}
\mathcal M(z,\eta)
={}&
32c_{1}(1+\eta^{2})\mathcal A
\\
&
+
c_{3}
\left[
21
-
9\eta^{2}
-
6z
-
8\eta^{2}z
-
2\eta^{4}z
\right].
\end{aligned}
\end{equation}

The charge shift is
\begin{equation}
Q_{1}^{2}
=
\frac{
32\mathcal A
}{
45(1-\eta^{2})^{2}
}
\mathcal Q(z,\eta),
\end{equation}
with
\begin{equation}
\begin{aligned}
\mathcal Q(z,\eta)
={}&
4c_{1}
\left(
5+10\eta^{2}+\eta^{4}
\right)
\mathcal A
\\
&
+
c_{3}
\left[
15
-
3\eta^{4}
-
20\eta^{2}z
+
8\eta^{4}z
-
4\eta^{6}z
\right].
\end{aligned}
\end{equation}

These expressions reduce smoothly to the corrected cold relations when \(\eta\rightarrow0\). In particular,
\begin{equation}
\lim_{\eta\rightarrow0}
M_{1}
=
\frac{
2(1-z)
}{
5r_{e}
}
\left[
32c_{1}(1-z)
+
c_{3}(7-2z)
\right],
\end{equation}

and
\begin{equation}
\lim_{\eta\rightarrow0}
Q_{1}^{2}
=
32(1-z)
\left[
c_{3}
+
4c_{1}(1-z)
\right].
\end{equation}

Thus the finite-distance family joins the corrected cold branch continuously.

Once \(M\) and \(Q\) are fixed, the cosmological horizon \(r_{c}\) follows as the largest positive root of the corrected radial function. We determine \(r_{c}\) perturbatively or numerically from
\begin{equation}
f(r_{c})=0,
\qquad
r_{c}>r_{+}.
\end{equation}

We use the resulting root when evaluating \(\kappa_{c}\) and imposing the outer boundary condition.

\subsubsection{Regular radial field}
The scalar equation derived previously can be written directly in the areal-radius coordinate by defining
\begin{equation}
P(r)
=
e^{\psi(r)}f(r).
\end{equation}

It then takes the form
\begin{equation}
P^{2}\Psi''
+
PP'\Psi'
+
\left(
\omega^{2}-V_{\ell}
\right)
\Psi
=
0.
\label{7.14}
\end{equation}

The quasinormal behavior at the two outer horizons is known analytically. We therefore factor it before discretizing the radial equation,
\begin{equation}
\Psi(r)
=
e^{\omega S(r)}
y(r),
\label{7.15}
\end{equation}
where
\begin{equation}
S(r)
=
-\frac{
i
}{
2\kappa_{+}
}
\ln(r-r_{+})
-
\frac{
i
}{
2\kappa_{c}
}
\ln(r_{c}-r).
\label{7.16}
\end{equation}

The remaining function \(y(r)\) is regular on the closed interval \(r_{+}\leq r\leq r_{c}\).

Substituting Eq. \eqref{7.15} into Eq. \eqref{7.14} and dividing once by \(P\) yields
\begin{equation}
\left[
\mathcal L_{0}
+
\omega\mathcal L_{1}
+
\omega^{2}\mathcal L_{2}
\right]
y
=
0.
\label{7.17}
\end{equation}

The frequency-independent operator is
\begin{equation}
\mathcal L_{0}y
=
P y''
+
P'y'
-
\frac{
V_{\ell}
}{
P
}
y.
\label{7.18}
\end{equation}

The term linear in frequency is
\begin{equation}
\mathcal L_{1}y
=
2PS'y'
+
\left(
PS''
+
P'S'
\right)
y,
\end{equation}

while the quadratic term is
\begin{equation}
\mathcal L_{2}y
=
\left[
P(S')^{2}
+
\frac{1}{P}
\right]
y.
\label{7.20}
\end{equation}

The apparent singularities in Eqs. \eqref{7.18}--\eqref{7.20} cancel once the exact horizon powers in Eq. \eqref{7.16} are included. The resulting equation is regular at both endpoints.

\subsubsection{Endpoint equations}
We impose the regular horizon limits analytically rather than evaluate expressions containing subtracting singular terms numerically.

Let \(r_{h}\) denote either \(r_{+}\) or \(r_{c}\), and write
\begin{equation}
q_{h}
=
P'(r_{h}),
\qquad
s_{h}
=
-\frac{
i
}{
2\kappa_{h}
}.
\end{equation}

Near the chosen horizon,
\begin{equation}
S'
=
\frac{
s_{h}
}{
r-r_{h}
}
+
S_{h}^{\mathrm{reg}}
+
\mathcal{O}(r-r_{h}),
\label{7.22}
\end{equation}

where \(S_{h}^{\mathrm{reg}}\) denotes the finite contribution from the opposite horizon factor. Since
\begin{equation}
s_{h}^{2}
=
-\frac{
1
}{
q_{h}^{2}
},
\end{equation}

the leading pole in \(\mathcal L_{2}\) cancels exactly.

We also define the finite horizon value
\begin{equation}
W_{h}
=
\left.
\frac{
V_{\ell}
}{
P
}
\right|_{r=r_{h}}.
\end{equation}

Taking the limit of Eq. \eqref{7.17} gives the regular endpoint equation
\begin{equation}
\begin{aligned}
0
={}&
\left[
q_{h}
+
2\omega q_{h}s_{h}
\right]
y'(r_{h})
\\
&
+
\Bigg[
-W_{h}
+
\omega
\left(
q_{h}S_{h}^{\mathrm{reg}}
+
\frac{
P''(r_{h})s_{h}
}{
2
}
\right)
\\
&
\qquad
+
\omega^{2}
\left(
2q_{h}s_{h}S_{h}^{\mathrm{reg}}
-
\frac{
P''(r_{h})
}{
q_{h}^{2}
}
\right)
\Bigg]
y(r_{h}).
\end{aligned}
\label{7.25}
\end{equation}

We obtained Eq. \eqref{7.25} by expanding the three operators around a simple horizon, and the divergent terms cancel.

The finite quantities appearing in Eq. \eqref{7.22} are especially simple,
\begin{equation}
S_{+}^{\mathrm{reg}}
=
\frac{
i
}{
2\kappa_{c}(r_{c}-r_{+})
},
\end{equation}

and
\begin{equation}
S_{c}^{\mathrm{reg}}
=
-\frac{
i
}{
2\kappa_{+}(r_{c}-r_{+})
}.
\end{equation}

Eq. \eqref{7.25} therefore supplies both endpoint rows of the spectral problem without introducing an auxiliary boundary prescription.

\subsubsection{Chebyshev discretization}
We map the exterior interval to
\begin{equation}
-1\leq\xi\leq1
\end{equation}

through
\begin{equation}
r(\xi)
=
\frac{
r_{c}+r_{+}
}{
2
}
+
\frac{
r_{c}-r_{+}
}{
2
}
\xi.
\end{equation}

For the pseudospectral discretization we use the standard Chebyshev-Lobatto grid \cite{Trefethen_2000},
\begin{equation}
\xi_{j}
=
\cos
\left(
\frac{
j\pi
}{
N
}
\right),
\qquad
j=0,\ldots,N.
\end{equation}

The first and second radial derivatives become finite matrices after the linear coordinate transformation. Evaluating Eqs. \eqref{7.18}--\eqref{7.20} on these nodes gives three matrices \(L_{0}\), \(L_{1}\), and \(L_{2}\). We replace their endpoint rows by Eq. \eqref{7.25}.

The resulting quadratic spectral equation has the standard quadratic-eigenvalue form \cite{Tisseur_2001},
\begin{equation}
\left(
L_{0}
+
\omega L_{1}
+
\omega^{2}L_{2}
\right)
\mathbf y
=
0.
\end{equation}

We use the usual companion linearization of this quadratic eigenvalue problem and introduce
\begin{equation}
\mathbf u
=
\begin{pmatrix}
\mathbf y
\\
\omega\mathbf y
\end{pmatrix}.
\end{equation}

The quasinormal frequencies then satisfy
\begin{equation}
\begin{pmatrix}
0&I
\\
-L_{0}&-L_{1}
\end{pmatrix}
\mathbf u
=
\omega
\begin{pmatrix}
I&0
\\
0&L_{2}
\end{pmatrix}
\mathbf u.
\end{equation}

This form permits us to obtain many frequencies at fixed \((z,\eta,\ell,c_{1},c_{3})\) in a single spectral solve.

\subsubsection{Mode identification and convergence}
We retain only eigenvalues that remain stable as the Chebyshev order increases, have regular reduced radial profiles, and move continuously under small changes of \(z\), \(\eta\), and the effective coefficients. Near cold extremality the \(\ell\geq1\) purely damped branches must approach Eq. \eqref{6.27}, while the fundamental \(\ell=0\) branch must approach Eq. \eqref{6.swavefreq}. The oscillatory large-\(\ell\) branch must approach the null-orbit result derived below. We make the numerical selection explicit by computing each candidate spectrum at \(N=40\), \(60\), and \(80\). For a candidate eigenvalue \(\omega_N\), we define the inter-order drift by \(\Delta_{N,N'}=|r_e\omega_N-r_e\omega_{N'}|/(1+|r_e\omega_N|)\). We also evaluate the normalized generalized-eigenvalue residual \(\rho_N=\|A_N\mathbf u-\omega_N B_N\mathbf u\|_2/[(\|A_N\|_2+|\omega_N|\|B_N\|_2)\|\mathbf u\|_2]\), where \(A_N\) and \(B_N\) are the two companion matrices. We discard a candidate as spurious unless \(\max(\Delta_{40,60},\Delta_{60,80})<10^{-7}\) and \(\rho_N<10^{-10}\) at the three orders. After normalizing the reduced radial vector by \(\max_j|y_j|=1\), we also require finite endpoint values and the absence of new grid-scale oscillations when \(N\) is increased. These requirements are applied before the spectral infimum is formed.

At fixed \(\ell\), we follow an accepted mode across parameter space by selecting the accepted eigenvalue closest in the complex-frequency plane to the preceding point. In regions where two accepted branches approach one another, we repeat the continuation from both directions and retain the separate branches. Family labels are then fixed by independent limiting behavior. The purely damped near-extremal family approaches the analytic cold spectrum, the oscillatory high-angular family approaches the null-orbit expression, and the de Sitter-type family approaches the pure-de Sitter frequencies given below. The finite-angular cascade discussed later is simpler because each competing branch has a distinct value of \(\ell\).

Three representative convergence sequences are shown below. The frequencies are reported as \(r_e\omega\). For an oscillatory conjugate pair we list the member with positive real part.

\begin{table}[t]
\caption{{\color{red}Representative Chebyshev-order convergence of accepted quasinormal modes. The residual column gives the conservative upper bound over \(N=40,60,80\).}}
\label{tab:qnm-convergence}
\centering
\resizebox{\columnwidth}{!}{%
\begin{tabular}{ccccc}
\hline
case & \(N=40\) & \(N=60\) & \(N=80\) & \(\max\rho_N\) \\
\hline
\(z=0.10,\ \eta=0.05,\ \ell=0\)
& \(-0.04000840893i\)
& \(-0.04000839127i\)
& \(-0.04000839125i\)
& \(<10^{-11}\) \\

\(z=0.10,\ \eta=0.13,\ \ell=4\)
& \(0.8953219855-0.06894974113i\)
& \(0.8953219820-0.06894973933i\)
& \(0.8953219820-0.06894973932i\)
& \(<10^{-11}\) \\

\(z=0.40,\ \eta=0.06300965,\ \ell=1\)
& \(0.05453564884-0.008344811393i\)
& \(0.05453564891-0.008344811469i\)
& \(0.05453564899-0.008344811498i\)
& \(<10^{-11}\) \\
\hline
\end{tabular}%
}
\end{table}

Table~\ref{tab:qnm-convergence} samples the near-extremal, photon-sphere, and finite-\(\ell\) regimes used in the family comparison. The largest visible inter-order change occurs for the near-cold \(\ell=0\) example between \(N=40\) and \(60\), while the \(N=60\) and \(80\) values agree to substantially higher precision. The two oscillatory examples show the same convergence pattern and remain well inside the acceptance thresholds.

We identify the de Sitter-type family independently by decreasing \(r_e\) at fixed \(\Lambda\) and fixed \(\eta\), with \(z=\Lambda r_e^2\). Along this path \(M,Q\rightarrow0\) and the exterior approaches the static de Sitter patch. The slower nonstationary scalar branch, which is the branch relevant to the spectral infimum, tends to the pure-de Sitter form \cite{Cardoso:2018nvb},
\begin{equation}
\omega_{n\ell}^{\mathrm{dS}}
=
-i\sqrt{\frac{\Lambda}{3}}
\left(\ell+2n\right),
\qquad
n=0,1,2,\ldots,
\end{equation}
with the \(\ell=n=0\) constant solution excluded from the decay rate. The second pure-de Sitter scalar branch is more strongly damped and does not affect the spectral-infimum comparison made here. We follow the fundamental nonstationary \(\ell=1\) member continuously back to the black hole of interest. This separates the outer family from purely imaginary near-extremal resonances and completes the family identification required for the spectral gap.

\subsection{Mode-family transitions}
\label{sec8}
The spectral gap changes physical origin along the charged de Sitter family. Near cold extremality the purely damped throat mode is slowest, while at larger normalized horizon separation an oscillatory photon-sphere family becomes dominant. Toward larger \(z\), finite angular momentum changes the ordering before the formal eikonal limit is reached. Since the gap is the lower envelope of the individual quasinormal branches, a change of the minimizing branch can also produce a kink in the parameter dependence of the gap even though each individual branch varies smoothly.

\subsubsection{Near-extremal and photon-sphere families}
The cold analysis gives the purely damped family directly. For the spherically symmetric fundamental mode, Eq. \eqref{6.swavefreq} gives the strict cold behavior. At \(z=0.1\) and \(\eta=0.05\) we find \(\kappa_-r_e=0.04468144\), \(\omega_{00}r_e\simeq-0.0400085i\), and \(\beta_{00}\simeq0.895\), consistent with the approach to \(\beta=1\).
The near-extremal dominance follows from the scale separation created by the nearly degenerate horizons. The throat modes inherit a damping scale set by the small near-horizon surface gravity, so their decay becomes slow as \(\eta\rightarrow0\). The photon-sphere and de Sitter-type families do not acquire the same cold suppression. The spectral gap is therefore supplied by the throat family sufficiently close to the cold endpoint.

For the oscillatory family we use the standard null-geodesic/eikonal correspondence \cite{Cardoso:2008bp} and define
\begin{equation}
A(r)=e^{2\psi(r)}f(r),
\qquad
U(r)=\frac{A(r)}{r^{2}},
\qquad
U'(r_{\mathrm{ph}})=0,
\qquad
U''(r_{\mathrm{ph}})<0.
\end{equation}
The orbital frequency and Lyapunov exponent are
\begin{equation}
\Omega_{\mathrm{ph}}
=
\frac{\sqrt{A(r_{\mathrm{ph}})}}{r_{\mathrm{ph}}},
\qquad
\lambda_{\mathrm{ph}}^{2}
=
-\frac{r_{\mathrm{ph}}^{2}f(r_{\mathrm{ph}})}{2}
U''(r_{\mathrm{ph}}).
\end{equation}
At large \(\ell\),
\begin{equation}
\omega_{\ell}^{\mathrm{ph}}
=
\ell\Omega_{\mathrm{ph}}
-\frac{i}{2}\lambda_{\mathrm{ph}}
+\mathcal O(\ell^{0}),
\qquad
\beta_{\mathrm{ph}}
=
\frac{\lambda_{\mathrm{ph}}}{2\kappa_-}
+\mathcal O(\ell^{-1}).
\end{equation}
This family depends on the exterior null orbit and therefore probes information not contained in the local extremal throat.
As the horizon separation grows, the cold suppression of the throat damping weakens, while the photon-sphere damping remains tied to the instability time of the outer unstable null orbit. The oscillatory family can then fall below the throat family. At larger \(z\), finite-wavelength corrections to the eikonal sequence become large enough to alter the angular ordering that determines the infimum. Our numerical spectrum establishes that ordering directly and shows that the minimum passes through a short sequence of finite \(\ell\) values.

\subsubsection{Transition and complete scalar gap}
At \(z=0.1\) and \(\eta=0.13\), the finite-angular ratios \(\beta_{\ell=1,2,3,4}\simeq0.51160,0.49713,0.49336,0.49182\) already show the exchange from the near-extremal branch to the photon-sphere sector. At the eikonal threshold, finite angular modes approach one half from above for \(z=0.2\) and \(0.3\). The corresponding values are \((0.51411,0.50465,0.50232,0.50139)\) and \((0.50265,0.50084,0.50042,0.50025)\) for \(\ell=1,2,3,4\). At \(z=0.4\) the ordering reverses. At the eikonal threshold we obtain \(\beta_{\ell=1,2,3}\simeq0.49830,0.49942,0.49971\), so the finite \(\ell=1\) mode crosses first.

\begin{figure}[t]
\centering
\includegraphics[width=0.60\textwidth]{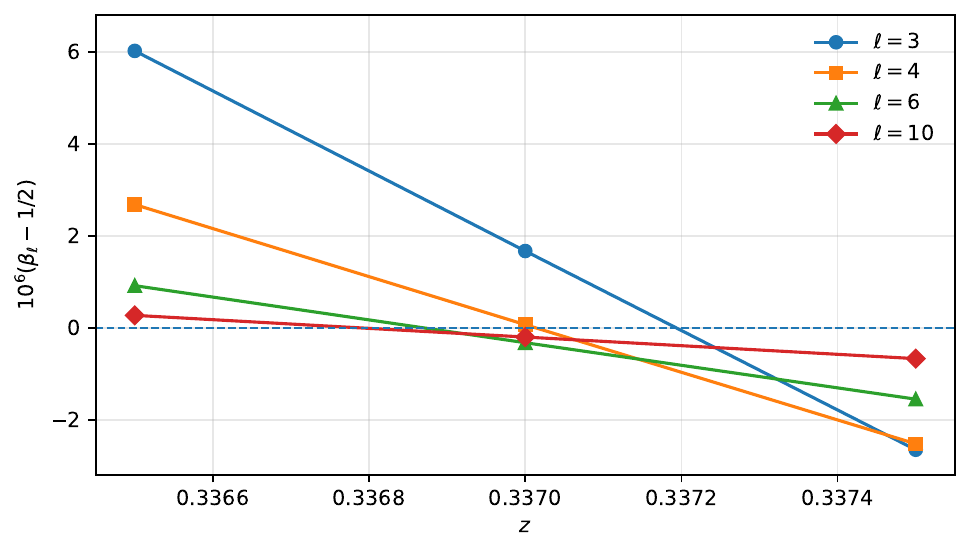}
\caption{Finite-angular departure from the eikonal neutral-scalar Christodoulou threshold. The plotted finite-\(\ell\) quantity is \(\beta_{\ell}-1/2\) evaluated at the two-derivative eikonal crossing for each \(z\). Negative values mean that the indicated finite-\(\ell\) mode reaches the threshold before the eikonal sequence. The first sign change occurs between \(z=0.3365\) and \(z=0.3370\). Above this onset the minimizing photon-sphere mode moves through successively lower angular sectors and is carried by \(\ell=1\) by \(z=0.3400\).}
\label{fig:angular-reordering}
\end{figure}

A denser sweep localizes the onset of this finite-angular reordering to
\begin{equation}
0.3365<z<0.3370.
\end{equation}
Figure~\ref{fig:angular-reordering} shows the sign change of the finite-angular excess over one half.

For the neutral scalar sector, following the Christodoulou spectral criterion for charged de Sitter black holes \cite{Cardoso:2017soq,Dias:2018etb}, the spectral gap is defined by
\begin{equation}
\alpha
=
\inf_{\substack{n,\ell\\ \omega_{n\ell}\neq0}}
\left\{-\operatorname{Im}\omega_{n\ell}\right\},
\qquad
\beta=\frac{\alpha}{\kappa_-}.
\end{equation}
Here \(\alpha\) denotes the exterior spectral gap. It is unrelated to the order-counting parameter \(\varepsilon\) used for the four-derivative expansion.
The infimum is required because the photon-sphere sequence can approach its limiting damping without attaining it at finite \(\ell\).
The physical meaning of the family change is then simple. When the throat family supplies \(\alpha\), the decay rate in the numerator and the inner-horizon blueshift scale in the denominator both inherit near-horizon physics. Once photon-sphere modes supply \(\alpha\), the numerator is fixed by the global exterior geometry while \(\kappa_-\) remains a Cauchy-horizon quantity. The Christodoulou ratio then compares scales associated with different regions of the spacetime. This change permits the finite-distance response to reverse sign even though the strict cold endpoint remains unchanged.

The lower-envelope character matters in the narrow angular-reordering region. A denser finite-angular calculation shows that the branch which reaches \(\beta=1/2\) first leaves the eikonal limit through large but finite \(\ell\). Representative points give \(\ell=6\) at \(z=0.3370\), \(\ell=4\) at \(z=0.3372\), \(\ell=3\) at \(z=0.3376\), and \(\ell=2\) at \(z=0.3382\). The \(\ell=2\) branch remains the first finite-angular crossing near \(z=0.339\), while \(\ell=1\) becomes the minimizing branch by \(z=0.3400\). The ordering is unchanged when the representative comparisons are repeated at spectral orders \(N=40,50,60\). Thus the transition is a short cascade of finite-angular branches rather than a direct differentiable continuation from \(\ell\rightarrow\infty\) to \(\ell=1\).

Table~\ref{tab:family-map} summarizes the parts of the \((z,\eta)\) plane that are directly fixed by the calculations above. It is intentionally restricted to the strict near-cold regime and to the two-derivative Christodoulou boundary. We do not infer a complete partition of the full two-parameter plane from these samples.

\begin{table}[t]
\caption{Representative source of the neutral-scalar spectral gap in the regions established directly in this work. The entries on the Christodoulou boundary refer to \(\eta=\eta_*^{(0)}(z)\). The eikonal entry denotes the large-\(\ell\) photon-sphere sequence. The angular-reordering band contains the finite-\(\ell\) cascade described in the text.}
\label{tab:family-map}
\centering
\begin{tabular}{ccc}
\hline
region & location & family supplying the gap \\
\hline
near-cold & \(\eta\rightarrow0\) & near-extremal throat, \(\ell=0\) \\
small and intermediate \(z\) boundary & \(z=0.05,0.10,0.20,0.30\) & eikonal photon sphere \\
reordering onset & \(0.3365<z<0.3370\) & eikonal to finite \(\ell\) \\
finite-angular cascade & \(0.3370\leq z<0.3400\) & finite-\(\ell\) photon sphere \\
larger-\(z\) boundary & \((z,\eta)=(0.40,0.06300965)\) & \(\ell=1\) photon sphere \\
\hline
\end{tabular}
\end{table}

We also compare the near-extremal and de Sitter-type branches at the representative neutral-scalar Christodoulou crossings. The near-extremal ratios at \(z=0.05,0.10,0.20,0.30,0.40\) are \(0.82399,0.82835,0.80731,0.67539,0.57594\). The continued de Sitter-type branch gives \(\beta_{\mathrm{dS}}\simeq0.81401,1.33093,2.75\) at the first three points and remains above \(4\) at \(z=0.30\) and \(0.40\). Thus neither family approaches the Christodoulou threshold at the crossings used below. The eikonal photon-sphere sequence supplies the infimum at the representative points through \(z=0.30\). After the narrow angular-reordering band, the finite \(\ell=1\) branch supplies the infimum from \(z=0.3400\) through the larger-\(z\) points considered here, including \(z=0.40\).

\section{Neutral-scalar Christodoulou boundary and coefficient-space interpretation}
\label{sec9}
\subsection{Displacement of the neutral-scalar Christodoulou boundary}
We now determine how the four-derivative interactions move the finite-distance Christodoulou boundary. Away from an exact mode-family exchange, let \(j\) denote the unique quasinormal branch that supplies the spectral infimum near the crossing. In the following equations we suppress the branch label and understand \(\beta\), \(b_i\), and \(B_i\) as quantities associated with that active branch. At two-derivative order we define \(\eta_*^{(0)}(z)\) by
\begin{equation}
\beta_0\left(z,\eta_*^{(0)}\right)=\frac12.
\end{equation}
At fixed \(r_e\), \(z\), and \(\eta\), write
\begin{equation}
\beta
=
\beta_0
+
\frac{\varepsilon}{r_e^2}
\left[b_1(z,\eta)c_1+b_3(z,\eta)c_3\right]
+
\mathcal O(\varepsilon^2),
\end{equation}
and
\begin{equation}
\eta_*(z)
=
\eta_*^{(0)}(z)
+
\frac{\varepsilon}{r_e^2}
\left[B_1(z)c_1+B_3(z)c_3\right]
+
\mathcal O(\varepsilon^2).
\label{9.3}
\end{equation}
Expanding the threshold condition gives
\begin{equation}
B_i(z)
=
-\frac{b_i\left(z,\eta_*^{(0)}\right)}{\left.\partial_\eta\beta_0\right|_*},
\qquad i\in\{1,3\}.
\end{equation}
The coefficient \(B_i\) therefore measures a displacement of the threshold in normalized horizon separation. It is distinct from the fixed-geometry correction \(b_i\) to the regularity ratio itself.
This derivative exists as an ordinary branch derivative when the minimizing mode is unique. For the monotonic threshold crossings considered here, at an exact exchange of two branches \(a\) and \(b\), the physical threshold is the lower envelope \(\eta_*=\min(\eta_*^{(a)},\eta_*^{(b)})\). If their response vectors differ, \(\mathbf B^{(a)}\neq\mathbf B^{(b)}\), the lower envelope has no single two-sided derivative at the exchange. Along any specified coefficient direction one instead uses the one-sided response of the branch that gives the smaller displaced threshold. We therefore treat \(B_i\) branchwise in the transition region and do not infer a unique response vector exactly at a mode crossing.
Since \(r_e=(r_++r_-)/2\) and \(\eta=(r_+-r_-)/(r_++r_-)\), the sign of \(\delta\eta_*\) states whether the threshold moves farther from or closer to the cold endpoint in normalized areal horizon separation. It does not define a volume on the full black-hole parameter space.

Table~\ref{tab:scc-boundary} summarizes the representative crossings. At \(z=0.1\), for example, \(\partial_\eta\beta_0|_*=-5.01440741\), \(b_1=30.86105181\), and \(b_3=4.28413287\), which gives \(B_1=6.15447635\) and \(B_3=0.85436474\). At \(z=0.40\), the finite \(\ell=1\) branch gives \(\partial_\eta\beta_0|_*\simeq-10.2140\), \(b_1\simeq-12.5563\), and \(b_3\simeq-4.18944\). Symmetric coefficient steps between \(10^{-4}\) and \(2.5\times10^{-5}\) in \(c_i/r_e^2\) change the final \(B_i\) values by less than \(10^{-5}\), while the Chebyshev-order variation gives the conservative uncertainties quoted in the table.

\begin{table}[t]
\caption{Two-derivative neutral-scalar Christodoulou crossings and first-order branch responses defined by Eq. \eqref{9.3}. Positive \(B_i\) moves the threshold to larger \(\eta\) for positive \(c_i\), while negative \(B_i\) moves it toward the cold endpoint. The first four rows lie in the eikonal photon-sphere region and the last row lies in the finite \(\ell=1\) region. None of the tabulated points is an exact mode-family exchange.}
\label{tab:scc-boundary}
\centering

\begin{tabular}{ccccc}
\hline
\(z\) & \(\eta_{*}^{(0)}\) & \(B_{1}(z)\) & \(B_{3}(z)\) & spectral family \\
\hline
0.05 & 0.13166844 & 6.82389 & 0.93384 & \(\ell\rightarrow\infty\) \\
0.10 & 0.12787994 & 6.15448 & 0.85436 & \(\ell\rightarrow\infty\) \\
0.20 & 0.11656132 & 4.50290 & 0.63001 & \(\ell\rightarrow\infty\) \\
0.30 & 0.09723997 & 2.20645 & 0.25718 & \(\ell\rightarrow\infty\) \\
0.40 & 0.06300965 & \(-1.2293(1)\) & \(-0.41016(3)\) & \(\ell=1\) \\
\hline
\end{tabular}

\end{table}

\begin{figure}[t]
\centering
\includegraphics[width=0.60\textwidth]{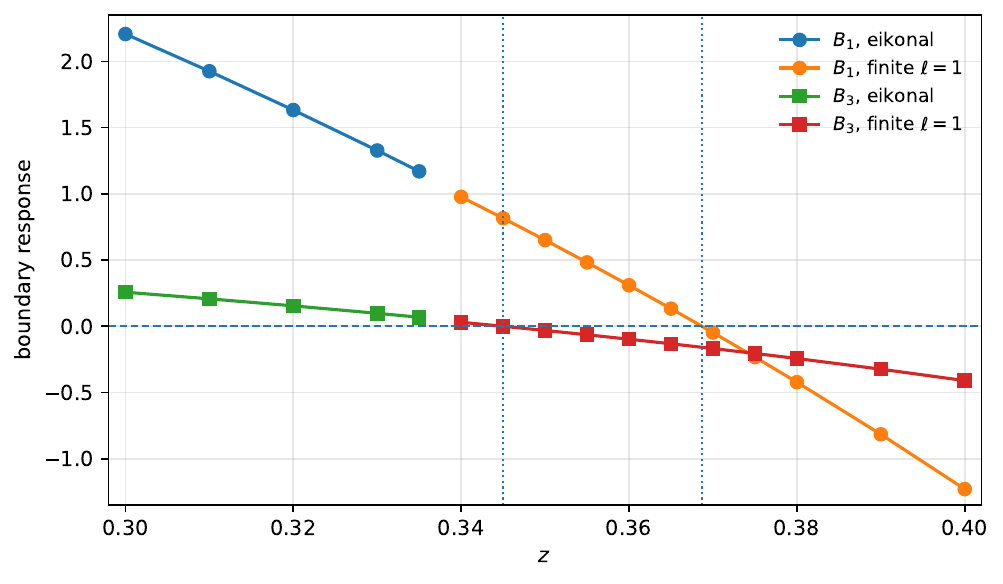}
\caption{Branchwise finite-distance neutral-scalar boundary response. The plotted functions are \(B_1(z)\) and \(B_3(z)\) from Eq. \eqref{9.3}. Positive values move the threshold to larger normalized horizon separation for a positive coefficient, while negative values move it toward the cold endpoint. The eikonal and finite \(\ell=1\) branches are shown explicitly. Intermediate finite-angular branches supply the lower envelope in the narrow family-exchange band, where a single derivative of the spectral gap need not exist. The dotted lines mark \(z_{B_3}=0.34497\) and \(z_{B_1}=0.36870\), both in the region where \(\ell=1\) uniquely supplies the gap.}
\label{fig:boundary-response}
\end{figure}

Positive \(c_1\) or \(c_3\) therefore moves the threshold to larger \(\eta\) at smaller and intermediate \(z\), but toward the cold endpoint at larger \(z\). The response must be read branchwise across the narrow family-exchange region and need not be differentiable at the exact exchange points. Once the \(\ell=1\) branch becomes the unique minimizer by \(z=0.3400\), its response components vary smoothly and give
\begin{equation}
z_{B_3}=0.34497,
\qquad
z_{B_1}=0.36870.
\end{equation}
These zeros are not located in the ambiguous exchange band. At \(z=0.34497\), the two-derivative crossing occurs at \(\eta_*^{(\ell=1)}=0.08431487\), while the \(\ell=2\) and eikonal crossings occur at \(0.08433719\) and \(0.08435291\), respectively. At \(z=0.36870\), the corresponding values are \(0.07592823\), \(0.07601924\), and \(0.07606750\). The \(\ell=1\) branch therefore supplies the spectral infimum uniquely at both zero locations. Repeating the coefficient-step and spectral-order tests leaves the rounded values \(z_{B_3}=0.34497\) and \(z_{B_1}=0.36870\) unchanged. Figure~\ref{fig:boundary-response} displays the eikonal and \(\ell=1\) branch responses on the two sides of the narrow exchange region.

The sign reversal does not conflict with the unchanged cold endpoint established by the normalized \(\ell=0\) solution. The dominant throat mode has \(\delta\beta_{00}|_{\eta=0}=0\) through first order, whereas \(B_i\) measures a finite-\(\eta\) crossing that depends on the relevant exterior scalar spectrum and unequal horizon surface gravities. Away from exact branch exchanges, at fixed \(z\), the first-order threshold is unchanged along
\begin{equation}B_1(z)c_1+B_3(z)c_3=0.
\end{equation}
No single coefficient combination therefore determines the boundary motion throughout the cold branch.

\subsection{Effective coefficient plane}
\label{sec10}
The preceding results define three linear responses in the \((c_1,c_3)\) plane. Using Eqs.~\eqref{3.13}, \eqref{3.23}, and \eqref{9.3}, we write
\begin{equation}
\begin{aligned}
\mathcal C_M(z) &= 8(1-z)c_1+(3+2z)c_3,\\
\mathcal C_A(z) &= 8(1-z)c_1+c_3,\\
\mathcal C_\beta(z) &= B_1(z)c_1+B_3(z)c_3.
\end{aligned}
\end{equation}
They govern different derivatives. The first fixes the fixed-charge extremal mass-parameter shift, the second fixes the local throat deformation and higher-angular cold spectrum, and the third fixes the finite-distance displacement \(\delta\eta_*=(\varepsilon/r_e^2)\mathcal C_\beta\). The form \(\mathcal C_\beta\) is branchwise whenever the spectral infimum changes family. It is well defined as a single linear form where one branch uniquely supplies the gap, including the two zero locations quoted above, but no unique \(\mathcal C_\beta\) exists at an exact exchange if the competing branch responses differ. Their zero lines therefore compare inequivalent physical responses in a common coefficient space.

The analytic zero-line slopes are
\begin{equation}
\left.\frac{c_3}{c_1}\right|_{\mathcal C_M=0}
=-\frac{8(1-z)}{3+2z},
\qquad
\left.\frac{c_3}{c_1}\right|_{\mathcal C_A=0}
=-8(1-z),
\qquad
\left.\frac{c_3}{c_1}\right|_{\mathcal C_\beta=0}
=-\frac{B_1}{B_3},
\end{equation}
when \(B_3\neq0\). The first two are distinct because \(\mathcal C_M-\mathcal C_A=2(1+z)c_3\), while the third is fixed only after solving the exterior spectral problem.

At \(z=0.1\), the three slopes are approximately \(-2.25\), \(-7.2\), and \(-7.20357\) for \(\mathcal C_M\), \(\mathcal C_A\), and \(\mathcal C_\beta\). The near coincidence of the latter two is accidental because they arise from different calculations. At \(z=0.4\), the slopes become \(-1.26316\), \(-4.8\), and \(-2.99716\). In addition, \(\mathbf B\) has reversed orientation, so positive \(c_1\) and positive \(c_3\) give \(\mathcal C_\beta<0\) there.

The sign structure makes the physical inequivalence explicit. A ray with \(\mathcal C_M>0\) decreases the extremal mass parameter, while the same ray can have either sign of \(\mathcal C_\beta\) as \(z\) changes. For \(\ell\geq1\), the strict cold spectral correction instead has the sign of \(\mathcal C_A\). A single coefficient ray can therefore decrease the extremal mass parameter, increase the higher-angular cold value of \(\beta\), and still move the finite-distance threshold toward the cold endpoint. Figure~\ref{fig:boundary-response} displays the change of the branch responses with cosmological scale. The physical lower envelope passes through the finite-angular cascade described in Section~\ref{sec8}, so we do not interpret the plotted eikonal and \(\ell=1\) curves as one differentiable response through the exchange band.

Thus no universal four-derivative direction governs all observables considered here. Extremality probes the cold charge-energy relation, the throat probes local geometry, and the neutral-scalar boundary response combines exterior decay with Cauchy-horizon amplification.

\subsection{Extremality and quantum-gravity interpretation}
\label{sec11}
The fixed-charge cold-extremal mass-parameter shift derived in Eq. \eqref{3.14} has the sign opposite to \(\mathcal C_M\). Hence \(\mathcal C_M>0\) decreases the extremal mass parameter throughout the ordinary cold branch. In the asymptotically flat limit the same direction becomes \(\mathcal C_M(0)=8c_1+3c_3\), where the \(1/r\) coefficient has its usual mass interpretation. At finite positive \(\Lambda\), we retain the more cautious language of a mass parameter because de Sitter spacetime has no unique analogue of the asymptotically flat ADM mass.

Away from an exact mode-family exchange, the neutral-scalar boundary response is governed instead by the branchwise form \(\mathcal C_\beta\). The distinction can be seen without tuning the coefficient ray. Along \(c_1>0\), \(c_3=0\), the extremal mass parameter decreases for every \(0\leq z<1/2\), while \(\delta\eta_*\) changes from \(+6.15448\,\varepsilon c_1/r_e^2\) at \(z=0.1\) to \(-1.2293\,\varepsilon c_1/r_e^2\) at \(z=0.4\). Along \(c_1=0\), \(c_3>0\), the corresponding shifts are \(+0.85436\,\varepsilon c_3/r_e^2\) and \(-0.41016\,\varepsilon c_3/r_e^2\). Therefore
\begin{equation}
\left.\delta M_{\mathrm{ext}}\right|_Q<0
\quad\not\Longrightarrow\quad
\operatorname{sgn}(\delta\eta_*).
\end{equation}
The failure follows from what the derivatives measure. The mass-parameter shift compares nearby degenerate solutions at fixed charge, while the neutral-scalar Christodoulou boundary compares the full exterior spectral infimum with the inner-horizon blueshift scale. The local throat introduces the third direction \(\mathcal C_A\). The three forms can coincide accidentally at isolated points, but they are generically inequivalent.

The weak-gravity comparison is therefore conditional. If an independently specified ultraviolet criterion and a chosen de Sitter mass prescription select coefficient directions corresponding to \(\mathcal C_M>0\), our result determines the neutral-scalar Christodoulou-threshold motion only after their location relative to the \(z\)-dependent line \(\mathcal C_\beta=0\) is known. At smaller \(z\), regions with \(\mathcal C_M>0\) can also have \(\mathcal C_\beta>0\). At larger \(z\), the same extremality sign can coexist with \(\mathcal C_\beta<0\). Extremality alone therefore cannot predict the fate of the neutral-scalar Cauchy-horizon regularity threshold.

Positive \(\Lambda\) is essential to this conclusion. It changes the cold throat ratio, introduces the cosmological horizon as the outer boundary of the wave problem, and reorganizes the competing quasinormal families. The local curvature-Maxwell contribution survives as \(\Lambda\rightarrow0\), but the finite-distance sign reversal depends on the global de Sitter exterior and would be absent from a calculation confined to the asymptotically flat limit.

We make no stronger claim that one unique de Sitter weak-gravity principle imposes a definite inequality on \((c_1,c_3)\). Different de Sitter extremal limits and mass prescriptions require additional assumptions. Our result instead provides the infrared map from any independently selected coefficient region to the direction in which the linear neutral-scalar Christodoulou threshold moves.

\section{Validity and scope}
\label{sec12}
\subsection{Perturbative domain and limits}
The four-derivative coefficients have dimensions of length squared, so we define
\begin{equation}
\lambda_4
=
\frac{\varepsilon}{r_e^2}
\max\left(|c_1|,|c_3|\right),
\qquad
\lambda_4\ll1.
\end{equation}
On the ordinary cold branch \(0\leq z<1/2\), this is a conservative local effective-theory condition. The cold throat adds the enhanced requirement
\begin{equation}
\left|
\frac{8\varepsilon(1-z)}{r_e^2(1-2z)}\mathcal C_A(z)
\right|\ll1,
\end{equation}
or, without relying on coefficient cancellations,
\begin{equation}
\frac{\varepsilon\max(|c_1|,|c_3|)}{r_e^2(1-2z)}\ll1.
\end{equation}
The immediate ultracold neighborhood therefore requires a separate scaling treatment.

The near-extremal parameter \(\eta\) is independent of the local effective expansion. If one perturbs two nearly coincident simple roots at fixed two-derivative \(M\) and \(Q\), Eq. \eqref{4.3} gives \(|\delta r_\pm|/r_e=\mathcal O(\lambda_4/\eta)\), so that particular expansion requires \(\lambda_4\ll\eta^2\). Our finite-distance parametrization instead fixes \(r_-\) and \(r_+\) geometrically and solves for \(M\) and \(Q\), which removes this artificial hierarchy. The analytic throat calculation keeps the leading \(\eta\) and \(\varepsilon\eta\) terms, while the finite-distance neutral-scalar Christodoulou boundary comes from the full exterior scalar problem.

The limits \(\eta\rightarrow0\) and \(\varepsilon\rightarrow0\) commute for the strict first-order value of the fundamental scalar endpoint because Eqs. \eqref{6.swavefreq} and \eqref{6.30} contain no \(\mathcal O(\varepsilon)\) term independent of \(\eta\). The limit \(\Lambda\rightarrow0\) is different. Local quantities have smooth limits, but the cosmological horizon moves to infinity and the global de Sitter quasinormal problem changes character. We therefore do not interpret the \(z\rightarrow0\) continuation of the finite-distance threshold as an asymptotically flat neutral-scalar Christodoulou boundary.

\subsection{Normalization, field variables, and spectral reliability}
A constant rescaling \(t' = e^C t\) sends \(\omega' = e^{-C}\omega\) and \(\kappa_a'=e^{-C}\kappa_a\). Hence
\begin{equation}
\frac{-\operatorname{Im}\omega'}{\kappa_-'}
=
\frac{-\operatorname{Im}\omega}{\kappa_-},
\end{equation}
so the regularity ratio is independent of the additive normalization of \(\psi\).

The field-variable convention was fixed with the reduced action in Section \ref{sec2}. We do not assign frame-independent meaning to \(c_{1}\) or \(c_{3}\) separately. Instead, the calculation refers to the explicitly specified gravitational action together with the minimally coupled neutral scalar. A consistent field redefinition of all sectors gives an equivalent description order by order, while changing the gravitational variables and then imposing new minimal scalar coupling would define a different spectator theory.

The spectral gap is established by comparing the near-extremal, de Sitter-type, finite-angular photon-sphere, and eikonal families rather than by following one branch. The representative sequences satisfy the inter-order acceptance condition stated above. For the three cases in Table~\ref{tab:qnm-convergence}, the \(N=60\) and \(80\) values differ at the \(10^{-10}\) level or better, and the normalized generalized-eigenvalue residuals remain below \(10^{-11}\). The dense angular sweep localizes the first departure from the eikonal ordering to \(0.3365<z<0.3370\). Repeating the branch comparison at \(N=40,50,60\) gives the same finite-angular sequence near representative points, with \(\ell=6\) at \(z=0.3370\), \(\ell=4\) at \(0.3372\), \(\ell=3\) at \(0.3376\), \(\ell=2\) at \(0.3382\), and \(\ell=1\) by \(0.3400\). Exact exchanges between consecutive branches are treated as lower-envelope kinks rather than as points with a unique derivative. Symmetric coefficient steps from \(10^{-4}\) to \(2.5\times10^{-5}\) in \(c_i/r_e^2\), together with the spectral-order tests, leave the zero locations \(z_{B_3}=0.34497\) and \(z_{B_1}=0.36870\) unchanged at the quoted precision. Both zeros lie well inside the \(\ell=1\) region rather than at a family exchange. The acceptance thresholds and continuation procedure stated in the numerical section are used for all branches entering these comparisons. The convergence table gives representative frequencies from three distinct spectral regimes and makes the numerical precision entering the lower-envelope selection explicit.

\subsection{Regularity criterion at the Cauchy horizon}
The local Christodoulou regularity test in terms of the spectral ratio is standard for this problem \cite{Cardoso:2017soq,Dias:2018etb}. Near the Cauchy horizon, with \(x=r-r_-\), the singular branch of a separated mode behaves as
\begin{equation}
|\Phi_s|\sim |x|^\beta,
\qquad
\beta=\frac{\alpha}{\kappa_-}.
\end{equation}
Hence \(|\partial_r\Phi_s|^2\sim |x|^{2\beta-2}\), and local square integrability requires
\begin{equation}
\int_0^{\epsilon_0}x^{2\beta-2}\,dx<\infty
\quad\Longleftrightarrow\quad
\beta>\frac12.
\end{equation}
The equality is marginal and can depend on logarithmic terms. Our conclusion therefore concerns sufficiently regular linear neutral scalar perturbations. It does not establish a nonlinear theorem for the full Einstein-Maxwell effective theory, and coupled gravitational or electromagnetic sectors can alter the Cauchy-horizon behavior.

\section{Conclusions}
\label{sec13}
We have studied leading parity-even four-derivative corrections to the linear neutral-scalar Christodoulou regularity problem for charged de Sitter black holes. The analysis separates the cold extremality relation, the local extremal throat, and the finite-distance competition between exterior decay and Cauchy-horizon blueshift. These three questions probe different coefficient directions.

At the cold endpoint, the scalar problem reduces to the finite-temperature \(AdS_2\times S^2\) throat. The \(\ell=0\) associated Legendre normalization becomes singular at the fundamental resonance, but the normalized limit gives the nonzero mode \(R_{00}=(x^2-1)^{-1/2}\) with \(\widehat\omega_{00}=-i\). The common near-horizon time scale then cancels from the regularity ratio, so the dominant nonstationary spherically symmetric mode retains \(\beta=1\) through first order. Higher angular sectors still respond to the corrected throat through \(\mathcal C_A\). The absence of a first-order shift in the fundamental endpoint therefore does not imply the absence of effective-theory corrections to the near-extremal spectrum.

The finite-distance neutral-scalar Christodoulou boundary behaves differently because the relevant exterior scalar spectrum enters. The photon-sphere family replaces the throat family as the source of the spectral gap, and finite angular momentum becomes decisive for larger \(z\). The first departure from the eikonal ordering occurs within \(0.3365<z<0.3370\), after which the minimizing photon-sphere branch passes through a short sequence of decreasing angular momenta and reaches \(\ell=1\) by \(z=0.3400\). The lower envelope can therefore have kinks at the exact exchanges even though each branch is smooth. Positive \(c_1\) and \(c_3\) move the Christodoulou threshold to larger normalized horizon separation at smaller and intermediate \(z\), but toward the cold endpoint at larger \(z\). The two response components change sign at \(z_{B_3}=0.34497\) and \(z_{B_1}=0.36870\). Both zeros occur where the \(\ell=1\) branch uniquely supplies the spectral gap, so the mode-crossing nondifferentiability does not alter their quoted values. We make no derivative assignment at the exact exchanges inside the preceding finite-angular cascade. The sign reversal is therefore a global de Sitter spectral effect rather than a deformation of the unchanged cold endpoint.

The coefficient-plane comparison makes the main implication precise. The fixed-charge extremal mass parameter responds to \(\mathcal C_M\), the local throat responds to \(\mathcal C_A\), and the finite-distance threshold responds to the \(z\)-dependent form \(\mathcal C_\beta\). A coefficient direction that decreases the extremal mass parameter can move the neutral-scalar Christodoulou threshold in either direction as the cosmological scale changes. Weak-gravity-inspired restrictions on extremality therefore do not determine this neutral-scalar response without additional information about the allowed coefficient region and the de Sitter mass prescription.

These conclusions hold only where the four-derivative and ordinary cold expansions remain perturbative and only for the minimally coupled neutral spectator scalar. They do not establish strong cosmic censorship for the full Einstein-Maxwell effective theory. Coupled gravito-electromagnetic perturbations can activate \(c_2\) and can change the spectral infimum. Future work should address those sectors, rotation, and the ultracold scaling regime, where the degenerating throat and effective corrections must be treated together. Such extensions will determine whether the cosmological-scale dependence found here persists in the full perturbative spectrum.

\begin{acknowledgments}
R. P. would like to acknowledge networking support of the COST Action CA21106 - COSMIC WISPers in the Dark Universe: Theory, astrophysics and experiments (CosmicWISPers), the COST Action CA22113 - Fundamental challenges in theoretical physics (THEORY-CHALLENGES), the COST Action CA21136 - Addressing observational tensions in cosmology with systematics and fundamental physics (CosmoVerse), the COST Action CA23130 - Bridging high and low energies in search of quantum gravity (BridgeQG), and the COST Action CA23115 - Relativistic Quantum Information (RQI), under COST (European Cooperation in Science and Technology).
\end{acknowledgments}

\bibliography{ref}

\end{document}